\documentclass[prd,superscriptaddress,showpacs,floatfix,preprintnumbers, nofootinbib,hyperref]{revtex4}              

\usepackage{graphicx}
\usepackage{rotating}
\usepackage{epsfig}

\usepackage[lofdepth,lotdepth,caption=false]{subfig}
\usepackage{bm}
\usepackage{amsmath}
\usepackage{color}
\begin{document}
\title{Searching for non-diagonal Mass varying mechanism in the $\nu_{\mu}-\nu_{\tau}$ system}

\author{D. R. Gratieri}
\affiliation{High and Medium Energy Group\\
Instituto de F\'isica e Matem\'atica\\
Universidade Federal de Pelotas\\
Caixa Postal 354, CEP 96010-900, Pelotas, RS, Brazil
}
\affiliation{Instituto de Fisica Gleb Wataghin - UNICAMP, 13083-859, Campinas, SP, Brazil}
\author{O. L. G. Peres}
\affiliation{Instituto de F\'isica Gleb Wataghin - UNICAMP, 13083-859, Campinas, SP, Brazil}
\affiliation{Abdus Salam International Centre for Theoretical Physics, ICTP, I-34010, Trieste, Italy}

\begin{abstract}
We use atmospheric neutrino data and MINOS data to constrain
 the MaVaN (Mass Varying Neutrinos) mechanism. The MaVaN model was largely studied 
in cosmology scenarios and comes from the coupling of the neutrinos with a neutral scalar   depending on the  local matter density. For atmospheric neutrinos,   this new  interaction affects the neutrino propagation  inside the Earth, and as consequence, induces modifications in their oscillation pattern.  To perform  such  test for a non-standard oscillation mechanism with a non-diagonal neutrino coupling in the mass basis,  we analyze the angular distribution of atmospheric  neutrino events as seen by the Super-Kamiokande experiment for the events in the  Sub-GeV and multi-GeV range and muon  neutrinos (anti-neutrinos) in MINOS experiment.  From the combined analysis of these two sets of data we  obtain the best fit for $\Delta m^2_{32}=2.45\times 10^{-3}$~eV$^2$,   $\sin^{2} (\theta_{23})=0.42$ and MaVaN parameter $\alpha_{32}=0.28$ with  modest improvement, $\Delta\chi^2=1.8$,  over the standard oscillation scenario.  The combination of MINOS data and Super-Kamiokande data prefers small values of MaVaN parameter $\alpha_{32}<0.31$ at 90\% C. L..
\end{abstract}

\date{\today}

\pacs{14.60.St,14.60.Pq,95.85.Ry}
\maketitle

\section{\label{sec:intro}Introduction}

Due to the  observations of  cosmic microwave background (CMB) radiation~\cite{planck, planck-II}, large scale structure (LSS)~\cite{Tegmark:2006az}, and Ia Supernovae~\cite{Riess:1998cb,Perlmutter:1998np,Astier:2005qq}   we known that the universe is actually  in accelerated expansion.  A direct way to incorporate this accelerated expansion of universe into Einstein General Relativity Theory~\cite{Einstein} is to include the cosmological constant $\Lambda$.  The inclusion of this constant has the same effect of a non-zero vacuum energy density, $\rho_{\rm vac}$, in such way that the pressure $p_{\rm vac}$ and the density $\rho_{vac}$ are related by a state equation with the form $p_{\rm vac}=-\rho_{\rm vac}$.

On the other hand, the accelerated expansion of universe can also be described by adding to the universe content a homogeneous fluid with energy density $\rho_{\lambda}$, the so called {\it Dark Energy} ~\cite{sweinberg}. This fluid has positive energy density $\rho_{\rm DE}$, and negative pressure $p_{\rm DE}$, in such way that  $p_{\rm DE}< -\rho_{{\rm DE}}/3$. This pressure is then responsible for the accelerated universe expansion. It is a remarkable fact that $73\%$ of universe content must be in the Dark Energy form. Also,  Dark Energy would be uniformly distributed in all space, and so, its density is constant in all points and times. This is  in contrast with the time evolution of baryonic  matter density $\rho_{\rm BM}$, that diminishes due to the expansion of the universe.  It is called {\it The Cosmic Coincidence} to the fact that today,  the baryonic matter density,  $\rho_{\rm BM}$ has approximately the same value of Dark Energy density $\rho_{\rm DE}$~\cite{dolgov}. To compare different cosmological models, it is common to define the density parameter $\Omega$ that it is the ratio between the density $\rho$ with the critical density $\rho_{\rm c}$ of the Friedmann universe, $\Omega=\rho/\rho_{\rm c}$.  In this sense The Cosmic Coincidence implies in the equality $\Omega_{\Lambda}=\Omega_{\rm BM}$.  This coincidence can be viewed  as an indicative  of  the existence of  some dynamical effect that relates both scales of baryonic matter and of the dark energy density.

Nevertheless, among all the models in the literature that are devoted to explain the accelerated expansion of universe, there is a class of dynamical models that obtain the desired negative pressure due to the inclusion of a scalar field that is the responsible for the variations in the expected value of vacuum energy. As the neutrino squared mass difference, $\Delta m^{2}_{32} \equiv m_3^2-m_2^2 = 2.5 \times 10^{-3}$ eV$^{2}$, where $m_3$ and $m_2$ are the masses of third and first neutrino mass state,  is of same order of dark energy scale,  it is straightforward  to think  a model in which  the scalar field couple to neutrinos and hence, the total energy of the fluid can vary slowly as the neutrino density decreases~\cite{Fardon:2003eh}. In the {\it Mass Varying Neutrino} models  the inclusion of a scalar field  allow the coupling of neutrino and Dark Energy densities due to the non-standard neutrino-scalar coupling.  The main consequence of this coupling to the neutrino physics is the fact that now the neutrino masses depends on the medium density.  This field could couples neutrinos to the baryonic matter and also to the neutrino background~\cite{Kaplan:2004dq}. The consequences in cosmology of MaVaN's had been studied in last years~\cite{Kaplan:2004dq,Fardon:2003eh,Gu:2003er,Bi:2003yr,Afshordi:2005ym,Honda:2005tw}  and references therein. 

For the neutrino phenomenology  the consequences of  MaVaN's  were studied for the solar and atmospheric neutrinos~\cite{Barger:2005mn,Cirelli:2005sg,GonzalezGarcia:2005xu,Shiraishi:2006gg,Abe:2008zza,deHolanda:2008nn,RossiTorres:2010mz,mateus}. 
We will use the data from Super-Kamiokande (SK for now) experiment~\cite{Abe:2008zza}.  In the experiment SK, down-going neutrinos, the ones  that are produce in the atmosphere immediately over Super-Kamiokande detector,  reaches SK with the  the cosine of zenith angle $\cos \theta_{z}\rightarrow 1$ and travels approximately 20 km in the atmosphere. On the other hand, up-going neutrinos are produced in the opposite side of the planet and  crosses all the Earth before reaches SK in the  the cosine of zenith angle $\cos \theta_{z} \rightarrow -1$ direction. The so-called {\it up-down} asymmetry of neutrino events in SK is recognized  as the first experimental corroboration of  neutrino flavor oscillations. We explicit here that atmospheric neutrinos that arrives at SK from different directions travels different distances and   crosses regions of very distinct densities (see next section for details). This fact makes the angular distribution of events in SK a good place to looking for dependence of medium density   effects in the propagation of neutrinos. In fact, the SK  experiment reported no improvement of data fit due to the inclusion of  diagonal MaVaN's mechanism in the propagation of atmospheric neutrinos~\cite{Abe:2008zza}. For our knowledge there is no analysis made for non-diagonal MaVaN's mechanism for atmospheric neutrinos. 

 We  also  apply the MaVaN's model to describe the beam muon neutrinos and anti-neutrinos at  MINOS experiment~\cite{Adamson:2013whj}, where a muon neutrino  or muon anti-neutrino beam  with energy between  few hundred of MeV and few GeV.  The neutrino or anti-neutrino beam  travels a few hundred of kilometers inside Earth's crust before reach the far detector. By the comparison between the number of neutrino events in near and far detector MINOS collaboration has measure precisely  the standard neutrino  parameters $\Delta m^{2}_{32}$ and $\sin^{2} (2\theta_{23})$. See for example~\cite{Adamson:2013whj,MINOS:beam} and references therein. In MaVaN's context, the  main difference between MINOS beam and SK is that in the former neutrinos cross only the upper crust of Earth, that can be described by a constant matter density and the latter
cross a different non-constant densities. In this way the MaVaN's effective oscillation for MINOS, with a constant density and with SK with a variable density allow us to test the essential characteristic of MaVaN mechanism, the density dependence of the neutrino mass differences.

The paper is organized as follows. In  the Section~(\ref{Sec:model}) we  describe the framework to MaVaNs that we adopt and in Sec.~(\ref{oscilograms}) we show the changes in neutrino oscillations due to MaVaNs. Then in Sec. (\ref{sec:muons}) we compare  our results for the angular distribution of events in SK  for neutrino oscillations without and with MaVaNs. Also in Section~(\ref{sec:minos}) we show how MaVaN's mechanism affects the allowed range of oscillations parameters for MINOS experiment and in Section~(\ref{sec:chi2}) we perform the $\chi^{2}$ analysis of SK atmospheric neutrino combined with  MINOS beam  data, and the constrains to MaVaN's model that we obtain.  Conclusions are in  
Section~ (\ref{sec:conclusion}).

\section{\label{sec:frame}Framework of MaVaN model}
\label{Sec:model} 
In the neutrino mass-mixing formalism, the time evolution of neutrino flavor eigenstates is given  in terms of neutrino mixing, in which one has to describe the flavor eigenstate  and hence, the time evolution of atmospheric neutrinos, in the two neutrino flavor approximation,  is given by the evolution equation 
\begin{eqnarray}
i\dfrac{d}{dt}\left(\begin{array}{c}
\nu_{\mu}\\
\nu_{\tau}
\end{array}\right)=
\left[ U H_{{\rm mass}} U^{\dagger}\right]\left(\begin{array}{c}
\nu_{\mu}\\
\nu_{\tau}
\end{array}\right)~,
\label{Sch}
\end{eqnarray}
and the Hamiltonian assumes  the form, in the mass basis, 
\begin{eqnarray}
H_{{\rm mass}}= \dfrac{\Delta m^{2}_{32}}{2E_{\nu}} 
\left[
\begin{array}{cc}
0 & 0  \\
0 &  1\\
\end{array}  \right]~,
\label{mmatrix}
\end{eqnarray}
here $E_{\nu}$ is the neutrino energy, $\Delta m^{2}_{32}\equiv m^2_3-m^2_2$ is the square difference of mass eigenstates, and  $U$ is the mixing matrix
\begin{eqnarray}
U=\left(
\begin{array}{cc}
c _{23}   &  s _{23}  \\
-s _{23}  & c _{23}   \\
\end{array}
\right)~,
\label{U}
\end{eqnarray}
where $c_{23}=\cos \theta_{23}, s_{23}=\sin \theta_{23}$.   The muon neutrino survival probability  is 
\begin{equation}
P(\nu_{\mu}\to \nu_{\mu})=1-\sin^2 2\theta_{23} \sin^2 \left(\dfrac{\Delta m^2_{32}L}{4E_{\nu}}\right),
\label{standardoscillation} 
\end{equation}
where  $\Delta m^2_{32}\equiv m_3^2-m_2^2$ is the squared mass difference, L is the distance and $E_{\nu}$ is the neutrino energy.  The formalism that we adopt  to include MaVaNs in the neutrino propagation has the Standard Model of particles plus a light scalar field $(\phi)$ with mass $m_{S}$ that couples with neutrinos  $\nu_{i}$, i=1,2,3  and with fermion fields f=e,n,p. In such model, the modification due to MaVaN's is the introduction of a fermion density dependent term  in each one of matrix elements in the neutrino mass matrix, Eq.~(\ref{mmatrix}), and so, including MaVaN's,  the flavor neutrino evolution is described by the generalization of Eq.~(\ref{Sch}). Explicitly we have
\begin{eqnarray}
i\dfrac{d}{dt}\left[\begin{array}{c}
\nu_{\mu}\\
\nu_{\tau}
\end{array}\right]=
H^{{\rm flavor}}_{{\rm MaVaN}}
\left[\begin{array}{c}
\nu_{\mu}\\
\nu_{\tau}
\end{array}\right], 
\label{eqevo-drg}
\end{eqnarray}
where the modified  Hamiltonian in the MaVaN's framework\footnote{In MaVaN mechanism neutrino and anti-neutrinos have the same oscillation probability.} is 
 \begin{eqnarray}
H^{{\rm flavor}}_{{\rm MaVaN}}= U H^{{\rm mass}}_{{\rm MaVaN}} U^{\dagger}
\label{hamil-drg}
\end{eqnarray}
where the mixing matrix U is defined in Eq.~(\ref{U}) and 
the MaVaN Hamiltonian has the form
\begin{eqnarray}
 H^{{\rm mass}}_{{\rm MaVaN}}=\dfrac{\Delta m^{2}_{32}}{2E_{\nu}}
\left[
\begin{array}{cc}
\alpha^{2}_{2}g(\rho) & \alpha^{2}_{32}g(\rho) \\
\alpha^{2}_{32}g(\rho) & 1+\alpha^{2}_{3}g (\rho) 
\end{array}\right]~.
\label{Mevo-drg}
\end{eqnarray}
Here   $\alpha_2,\alpha_3$ and $\alpha_{32}$ are the MaVaN's parameters and  $g(\rho)$  is the function of the Earth matter density $\rho$ that neutrinos feels while cross the Earth.   When  $\alpha_2=\alpha_3=\alpha_{32}=0$ we recover the standard neutrino evolution given in Eq.~(\ref{Sch}). We can classify  the behavior of  MaVaN mechanism given in Eq.~(\ref{Mevo-drg}) in two types: (a) when $\alpha_2,\alpha_3 \neq 0$ and $\alpha_{32}=0$ and (b) when $\alpha_2=\alpha_3=0$ and $\alpha_{32}\neq 0$. In the former case the  neutrino survival probability, for constant density,   is given by  
\begin{equation}
P(\nu_{\mu}\to \nu_{\mu})=1-\sin^2 2\theta_{23} \sin^2 \left(\dfrac{(\Delta m^2)^{\prime}_{{\rm eff}}L}{4E_{\nu}}\right),
\label{case1MaVaN}
\end{equation}
where the effective mass scale  is given by $(\Delta m^2)^{\prime}_{\rm eff}\equiv \Delta m^{2}_{32} \left[1+(\alpha^{2}_{3}-\alpha^{2}_{2}) g(\rho)\right]$ . In this case, when $\alpha_{32}=0$  the amplitude of oscillations, $\sin^2 2\theta_{23} $ is the same as in the  standard neutrino oscillations in Eq.~(\ref{standardoscillation}), and the phase of the oscillations, proportional to $(\Delta m^2)^{\prime}_{\rm eff}$,  have now a matter density dependence.  This was the case mostly worked in the literature~\cite{Barger:2005mn,Cirelli:2005sg,GonzalezGarcia:2005xu,Shiraishi:2006gg,Abe:2008zza,deHolanda:2008nn,RossiTorres:2010mz,mateus}. The latter case,  the Hamiltonian in the mass basis is non-diagonal and for our knowledge it was {\it not explored} in the literature for atmospheric neutrino phenomenology. The probability, for constant density $\rho$,  can be written as 
\begin{equation}
P(\nu_{\mu}\to \nu_{\mu})=1- \sin^2  2\theta_{{\rm MaVaN}} \sin^2\left( \dfrac{\Delta m^2_{{\rm MaVaN}}L}{4E_{\nu}}\right),
\label{case2MaVaN}
\end{equation}
where the amplitude,  $\sin^2  2\theta_{{\rm MaVaN}}$ and the phase of oscillations, $\Delta m^2_{{\rm MaVaN}}$ are different from the usual standard oscillation scenario.  The MaVaN  mass difference  is given by
\begin{equation}
\Delta m^2_{{\rm MaVaN}}\equiv \Delta m^{2}_{32}  \sqrt{ \{2 \alpha^{2}_{32}g(\rho)\}^2 + 1}~.
\label{Eq:effDM}
\end{equation}
where the MaVaN  mass difference  depends on the medium density and the explicit expression for the amplitude is 
\begin{equation}
\sin^2  2\theta_{{\rm MaVaN}}\equiv \sin^2  (2\theta+2\eta) =\left[\sin  (2\theta)\cos (2 \eta)+\sin  (2\eta)\cos (2 \theta)\right]^2
\label{Eq:sineff}
\end{equation}
that also have a dependence on the medium density. The angle $\eta$ is defined as
\begin{equation}
\sin^{2}( 2\eta)= \dfrac{
\left[2\alpha^{2}_{32}g(\rho)\right]^2}
{1  +\left[ 2\alpha^{2}_{32}g(\rho)\right]^{2} }
\label{Eq:neta}
\end{equation}
in MaVaN's case for $\alpha^{2}_{32}\neq 0$ induce that the mass basis it is not diagonal and the parameter $\eta$ is the mixing angle that diagonalizes the mass basis as shown in reference~\cite{diegophdthesis}.  This made that the non-diagonal MaVaN's would induces neutrino oscillations even if the vacuum mixing angle was zero, $\theta \to 0$.

In the case of standard oscillations, see Eq.~(\ref{standardoscillation}) we have the  symmetry   $\sin^2 \theta_{23}\leftrightarrow \cos^2 \theta_{23}$, but in MaVaN case with $\alpha^{2}_{32}\neq 0$ we broken this symmetry and the   Eq.~(\ref{Eq:sineff}) have different results for $\sin^2 \theta_{23}>\cos^2 \theta_{23}$ compared when $\sin^2 \theta_{23}< \cos^2 \theta_{23}$.  For vanishing MaVaN parameters, $\alpha^{2}_{32}=0$, we have $\sin^2  2\eta \to 0\quad \eta \to 0$, and we recover the standard neutrino oscillation. For very large values of MaVaN parameters we have $\sin^2  2 \eta \to 1$, that implies that $\sin^2  2\theta_{{\rm MaVaN}}\to \cos^2  2\theta_{23}$. If we have large mixing angles $\theta_{23} \sim \pi/4$ we will have suppression of the amplitude, but for smaller mixing angles will have a enhancement of the amplitude.

Most of the previous analysis of MaVaN works for the first two generations~\cite{Barger:2005mn,Cirelli:2005sg,GonzalezGarcia:2005xu,deHolanda:2008nn,RossiTorres:2010mz,mateus} 
and then theirs constrains are related to parameters of the first generation $\alpha_1,\alpha_2$ and $\alpha_{12}$ similar to the parameters defined in Eq.~(\ref{Mevo-drg}). From the Reference~\cite{GonzalezGarcia:2005xu} that provide a upper bound for the elements of matrix $|H_{12}|_{\rm MaVaN},|H_1|_{\rm MaVaN}<10^{-4}$~eV at 90 \% C.L.  The only case that works in the MaVaN scenario for the second and third families  are  the References~\cite{Shiraishi:2006gg,Abe:2008zza}. In this works  they assume the non-zero parameters $\alpha_2$ and the medium matter dependence is given by $g(\rho)=(\rho/\rho_0)^n$, where $\rho$ is the matter density, $\rho_0=1{\rm g/cm}^3$  and {\it n} is a free parameter.  It was ruled out diagonal MaVaN to be the dominant oscillation scenario for atmospheric neutrinos and they constraint the  {\it n} parameter to be in the range $-0.15<{\it n}<0.1$ at 90 \% C.L.~\cite{Shiraishi:2006gg,Abe:2008zza}.

\subsection{Oscillation probabilities without MaVaN's and with MaVaN's}
\label{oscilograms}

Now we are going to compare the oscillation probabilities  without MaVaN's and with MaVaN's to understand the changes in the oscillation probabilities.  

Most of previous analysis use a model for the MaVaN's mass $M_i$ given as a function of parameter with dimensions of energy such as   $M_i=\mu_i g(\rho)$, with the index i denoting the diagonal mass eigenstates i=1,2 and the non-diagonal mass eigenstates i=3;  with different functions $g(\rho)$ as function of the matter density $\rho$,   and a  parameter $\mu_i$ with dimension of energy.  We decide to adopt a dimensionless parameters in this work, the $\alpha_i$ as described  in Eq.~(\ref{Mevo-drg}), but we can related our parameters $\alpha_i$ with the previous analysis by the replacing of   $\mu_i \to \alpha_i \Delta m^2$, where  $\Delta m^2$ is the relevant mass difference of the analysis.

\begin{figure}[t!]
\centering
\subfloat[]{
 \label{prob-cz-04}
\includegraphics[scale=0.33]{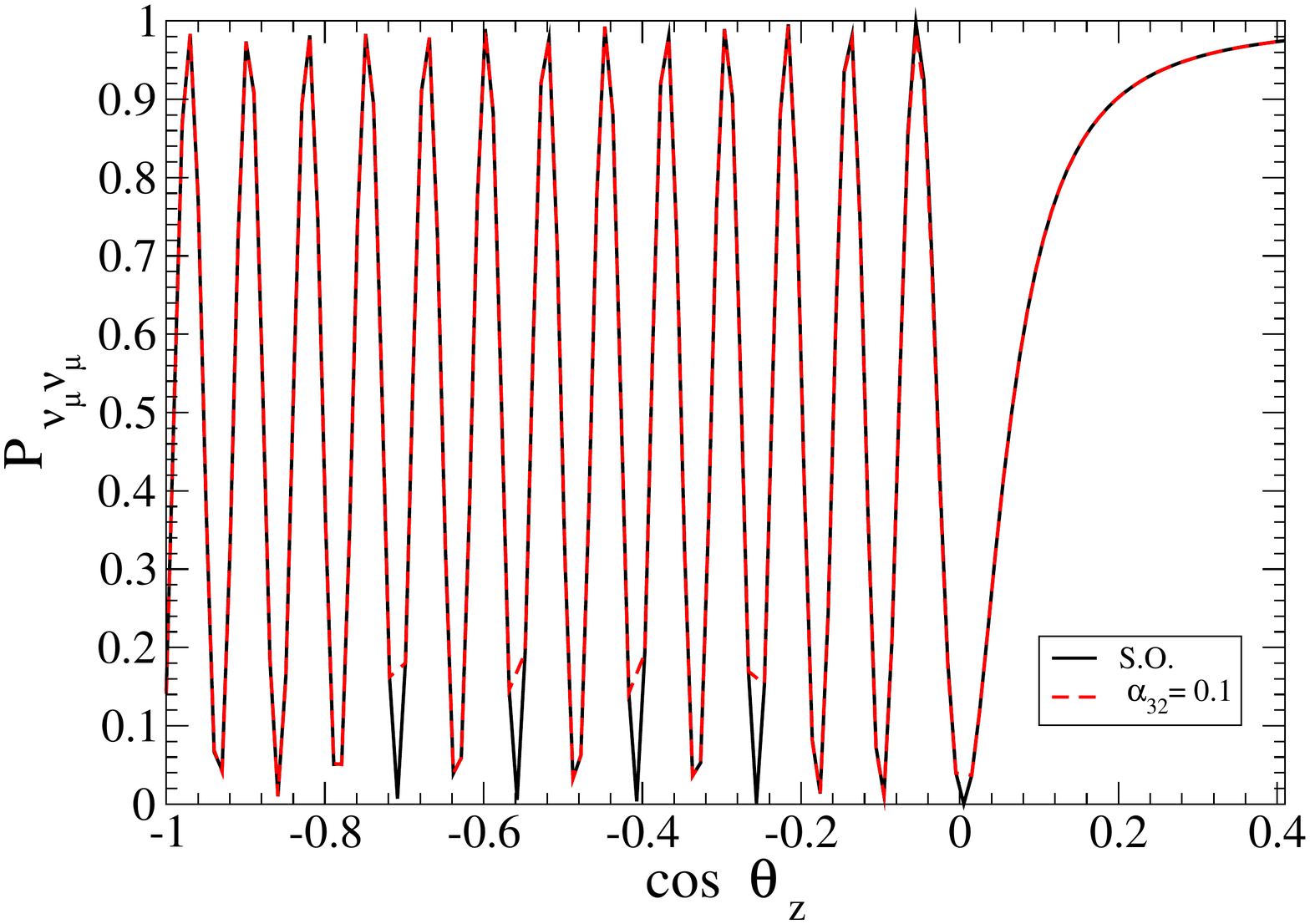}
\hspace{-3.42em}
}
\subfloat[]{
\includegraphics[scale=0.33]{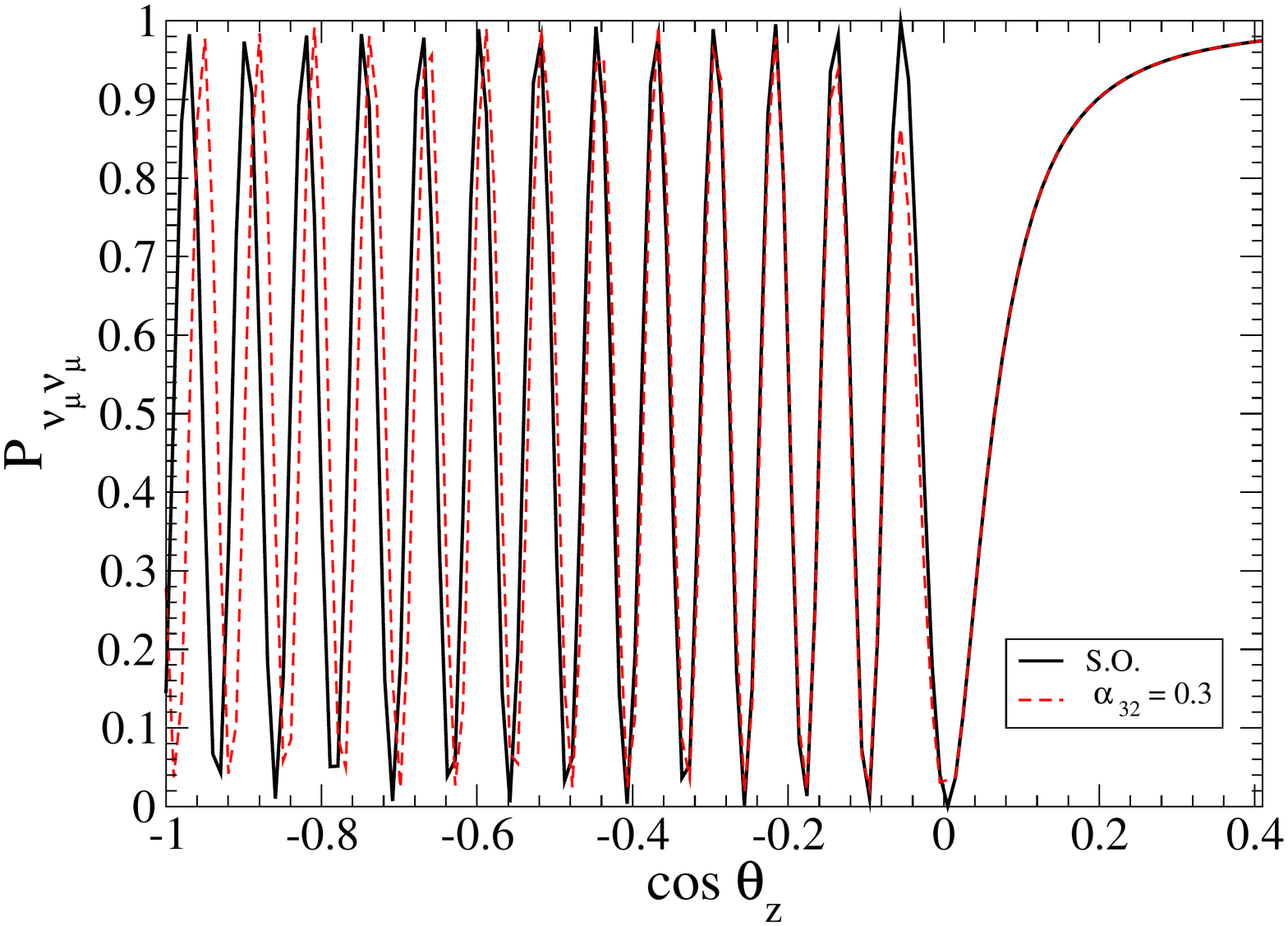}
  \label{prob-cz-01}
}
\quad
\subfloat[]{
\includegraphics[scale=0.33]{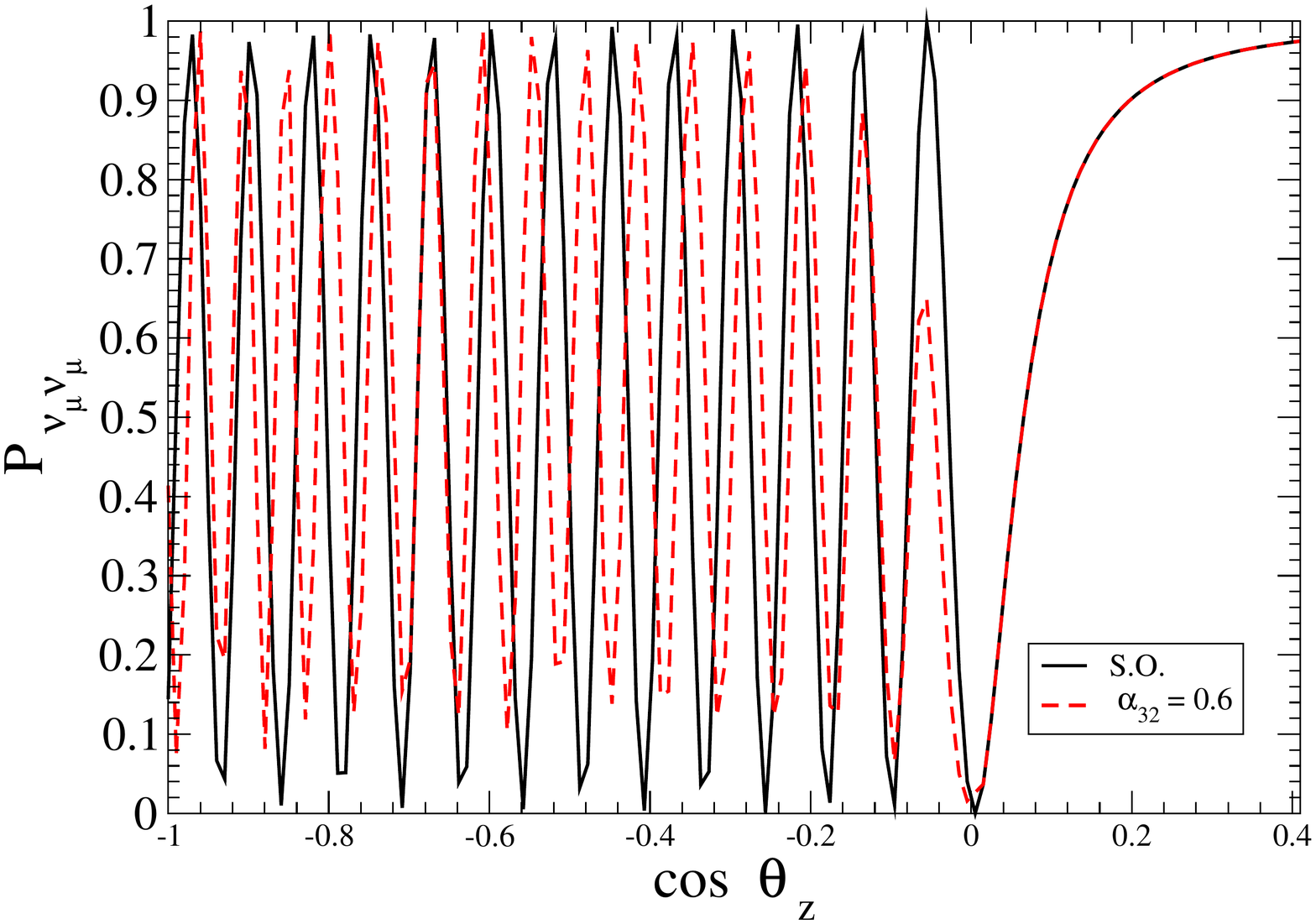}
  \label{prob-cz-02}
\hspace{-3.42em}
}
\subfloat[]{
\includegraphics[scale=0.33]{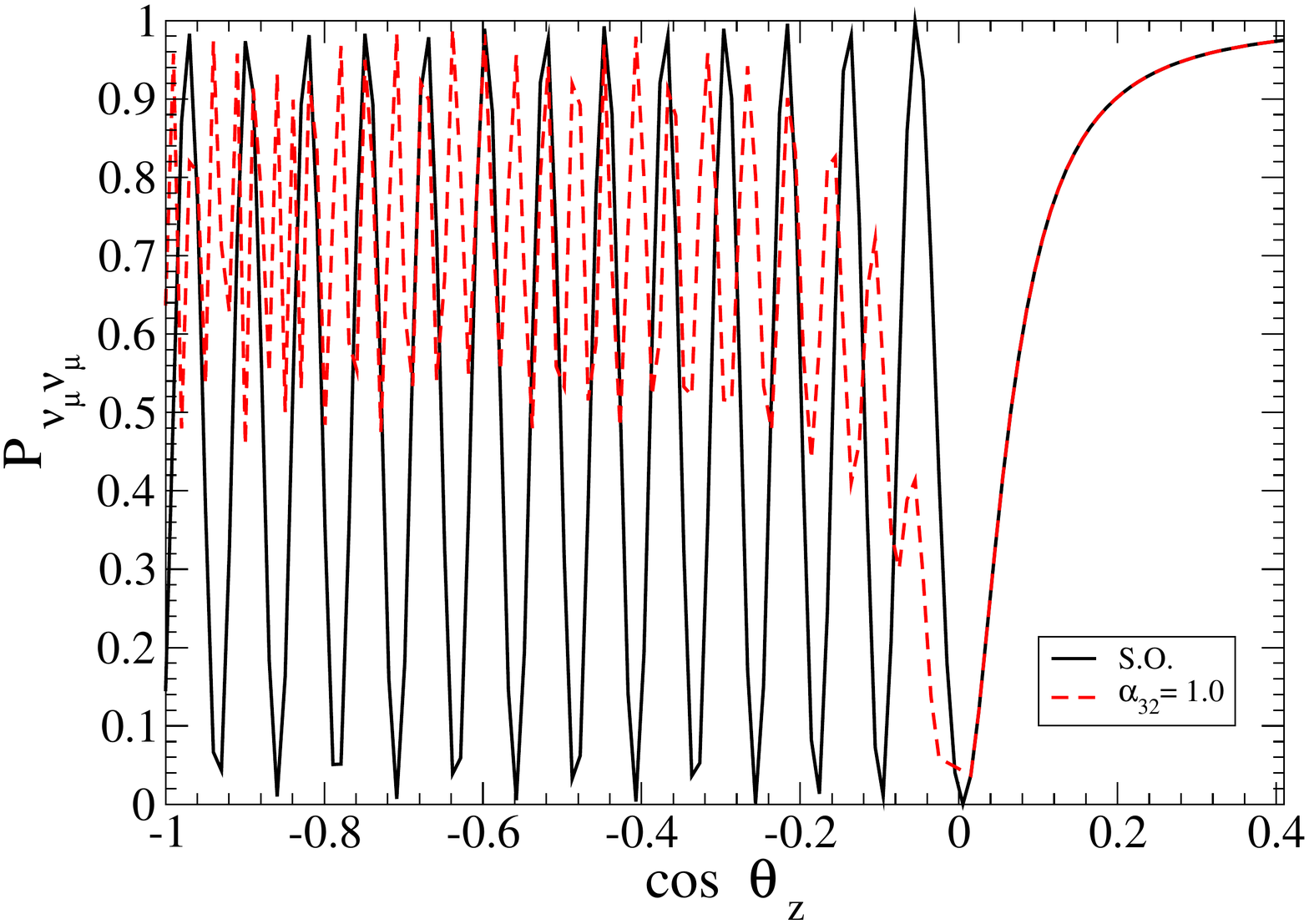}
  \label{prob-cz-03}
}
\caption{ Muon neutrino survival probability as function of $\cos \theta_{z}$.  The standard oscillation (S. O.) is shown in black solid curve and the MaVaN's oscillation curves are in dashed color  with the values of $\alpha_{32}$ are indicated in each panel. In this plots we assume the values of $\sin^{2}(2\theta_{23})=1.0$ and $\Delta m^{2}_{32}=2.6\times 10^{-3}$ eV$^{2}$ and $E_{\nu}=1.0$~GeV in all  plots.}
\label{fig:pmumu-cz}
\end{figure}

In this work we decide to study the case when for non-diagonal MaVaN parameter, $\alpha_2=\alpha_3=0$ and $\alpha_{32}\neq 0$, vide Eq.~(\ref{Mevo-drg}).
As an example e will use in this work the MaVaN density dependence as used in Ref.~\cite{deHolanda:2008nn,mateus}, given by
\begin{equation}
g(\rho)= tanh^{2} \left( \dfrac{\rho}{\rho_{\rm core}} \right)~,
\label{dmevo1}
\end{equation}
In this work we will use the matter profile of Earth that we take from ~\cite{PREM}.
where $\rho$ is the matter density that neutrino crosses and $\rho_{\rm core}=11.5$ g/cm$^{3}$ is the matter  density at {\rm Earth} core. This choice is motivate to generates a soft and well behaved function even with the abrupt variations of {\rm Earth} density profile, other  choices made the computation numerical unstable as reported in Ref.~\cite{Shiraishi:2006gg}. Also it have a finite limit for $\rho\to \infty$ and $g(\rho)<1$ always.  In this work we will use the matter profile of Earth that we take from~\cite{PREM}.

\begin{figure}[t!]
\centering
\subfloat[]{
\includegraphics[scale=0.6]{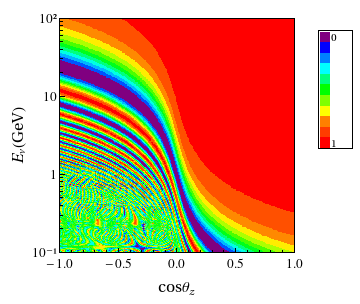}
  \label{oscillogram-1-MaVaN}
}
\subfloat[]{
\includegraphics[scale=0.6]{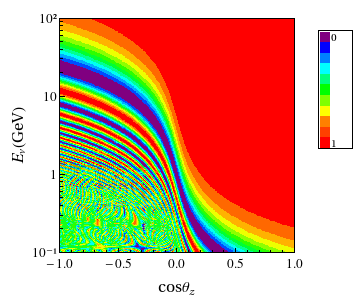}
  \label{oscillogram-2-MaVaN}
}
\quad
\subfloat[]{
\includegraphics[scale=0.6]{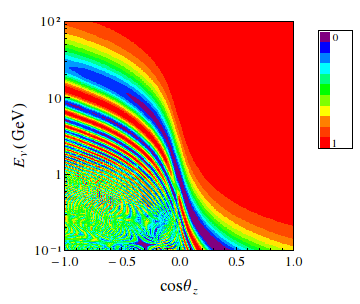}
  \label{oscillogram-3-MaVaN}
}
\subfloat[]{
  \includegraphics[height=6.5cm,width=8cm]{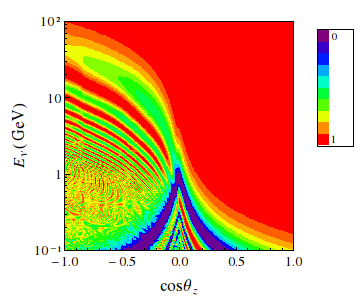}
  \label{oscillogram-4-MaVaN}
}
\caption{In sequence, upper-left, upper right, lower left and lower-right
we show the oscillogram in the plane cosine of neutrino zenith
($\cos\theta_{z}$) angle and neutrino energy ($E$), for the survival
probability $P(\nu_{\mu}\to\nu_{\mu})$ respectively for
$\alpha_{\rm{32}}=0.0,~0.1,~0.5,~1.0$. We assume $\Delta m^2_{32}=2.6\times 10^{-3}$eV$^2$ and maximal mixing angle $\sin^2(\theta_{23})=1/2$.}
\label{fig:pmumu_alpha0.1}
\end{figure}

We solve numerically compute the muon neutrino survival probability, $P(\nu_{\mu} \to \nu_{\mu})$  for the general case of a neutrino crossing the Earth from different chords. We can related the traveled distance by neutrino {\it L} to the zenith angle $\theta_z$ by $L = - R_{\rm E}cos(\theta_{z}) + \sqrt{R^{2}_{\rm E}cos^{2}(\theta_{z})+h^2+2 R_{\rm E}h}$, where $R_{{\rm E}}=6371$ km is {\rm Earth} radius and {\it h} is the point of atmosphere were neutrinos are produced, approximately $20$ km. For $\cos\theta_z\to 1 (-1)$ we have the maximum (mininum) distance. We shown in Fig.~(\ref{fig:pmumu-cz}) the muon neutrino probability  as a function of cosine of zenith angle, $\cos \theta_z$  for fixed values of the amplitude of the mixing angle $\sin^{2}(\theta_{23})=1/2$, for the mass difference $\Delta m^{2}_{32}=2.6\times 10^{-3}$ eV$^{2}$ and for a fixed energy $E_{\nu}=1.0$~GeV and several values of  $\alpha_{32}$: $\alpha_{32}={0.1, 0.3, 0.6, 1.0}$ we have respectively the MaVaN probabilities as dashed curves and without MaVaN as solid curves in  Fig.~(\ref{prob-cz-04}, (\ref{prob-cz-01}),(\ref{prob-cz-02}) and (\ref{prob-cz-03}).  For values of $\cos (\theta_{z})>0$, the neutrino traveled in vacuum only  and the oscillation probability with and without MaVaN are the same. For $\cos (\theta_{z})<0$, the neutrino  cross inside the Earth and the MaVaN effect begin to pop up.  To see more clearly the effect of MaVaN parameters we can  compare   the Fig.~(\ref{prob-cz-04}) that have $\alpha_{32}=0.1$, with Fig.~(\ref{prob-cz-03})  that have $\alpha_{32}=1.0$. When can see when we increase  the MaVaN parameter $\alpha_{32}$, for maximal mixing, the MaVaN amplitude get smaller and the oscillation phase increases giving  more fast wiggles that appear in the probability. The increase of wiggles made the maximums and minimums move to smaller values of  $\cos \theta_{z}$.

\subsection{Oscillograms}

The concept of oscillograms is interesting tool to understand the complete behavior of neutrino probability in some model for neutrino oscillation. We plot in Fig.~(\ref{fig:pmumu_alpha0.1}) the oscillograms of muon neutrino survival probability, denoted by $P_{\nu_{\mu}\nu_{\mu}}$,  as function of neutrino energy $E$ and the cosine of zenith angle, $\cos(\theta_{z})$.  In the Fig~(\ref{oscillogram-1-MaVaN}) we show for the standard neutrino case and in the others plots for increasing bigger values of MaVaN parameter.  For the cosine of zenith angle, $\cos(\theta_{z})>0$, when the neutrino did not cross the {\rm Earth}, we have zero MaVaN effect and for  $\cos(\theta_{z})<0$, the muon survival probability is modified due MaVaN mechanism. 
For comparison we show in the upper-left panel, in the upper-right panel,   lower-left panel  lower-right panel of n Fig.~(\ref{fig:pmumu_alpha0.1}) for MaVaN parameters $\alpha_{32}$ respectively equal to 0.0, 0.1, 0.5 and 1.0.   An enhancement of $\Delta m^{2}_{\rm MaVaN}>\Delta m^{2}_{32}$ also implies that the positions of maximums and minimums is dislocated to higher values of neutrino energy. As an example of this let we look to the first minimum in $P_{\nu_{\mu}\nu_{\mu}}$ for $\cos\theta_{z}=1.0$. In the former three panels of Fig~(\ref{oscillogram-1-MaVaN}) we see that the first minimum ocurs for $E_{\nu}\approx 25$~GeV. In the lower-left panel, where $\alpha_{32}=1.0$ this minimum in $P_{\nu_{\mu}\nu_{\mu}}$ had its intensity reduced (due to the increase of $\sin^{2}\theta_{\rm MaVaN})$ but also we see that the minimum was dislocated to $E_{\nu}\approx 40$~GeV. At $E_{\nu}\approx 25$~GeV now we see the first maximum of oscillation that for S.O. occur at $E_{\nu}\approx 4.5$~GeV. The same kind of displacement is found for all the maximums and minimums in the lower-left panel of  Fig~(\ref{oscillogram-1-MaVaN}) when compared with the cases in which $\alpha_{32}<1.0$.

\section{\label{sec:muons}Number of muons and electrons in Super-Kamiokande experiment}

Atmospheric neutrinos, composed muon neutrinos and electron neutrinos, are  produced all around the Earth atmosphere and traveled to Super-Kamiokande from all directions. Once in the detector they interact producing muons and electrons and Super-Kamiokande measure the zenith angle dependence of these muons and electrons.
The rate for these events can be computed as 
\begin{eqnarray}
N(\mu)&=&T N_{t}~
\int_{E^{\nu}_{0}}^{E^{\nu}_{f}}dE_{\nu}
\int_{0}^{1}dx
\int_{-1}^{1} d (cos \theta_{z})
\int_{0}^{2\pi} d \phi_{z}
\int_{cos\theta_{\mu_{0}}}^{\cos \theta_{\mu_{f}}} d\cos\theta_{\mu}~  
\nonumber \\
&\times& \left \{
\dfrac{d^3\Phi_{\nu_{\mu}}(E_{\nu},\theta_{z},\phi_{z})}{dE_{\nu}d(\cos\theta)d\phi_z}
\times P(\nu_{\mu}\rightarrow \nu_{\mu})\times 
\dfrac{d\sigma_{\nu_{\mu}}(E_{\nu},E_{\mu})}{dE_{\mu}} 
\right.  \nonumber \\
&+ & \left. 
\dfrac{d^3\Phi_{\bar{\nu}_{\mu}}(E_{\nu},\theta_{z},\phi_{z})}{dE_{\nu}d(\cos\theta)d\phi_z}
\times P(\bar \nu_{\mu}\rightarrow \bar  \nu_{\mu})\times
 \dfrac{d\sigma_{ \bar \nu_{\mu}}(E_{\nu},E_{\mu})}{dE_{\mu}} 
\right \} \nonumber \\
&\times& \Theta[E_{\mu}(\cos \theta_{z},E_{\nu},\cos\theta_{\mu})-E^{{\rm min}}_{\mu}]\Theta[E^{{\rm max}}_{\mu}-E_{\mu}(\cos\theta_{z},E_{\nu},\cos\theta_{\mu})]
 ~,
\label{nmu}
\end{eqnarray}
where $N_{t}$ is the number of targets in SK, $T$ is the livetime , $E_{\nu}$ is the neutrino energy, $\cos\theta_z$ is the cosine of zenith angle ($\theta_z$)   of the neutrino,
$\phi_z$ is the azimuth  angle of the neutrino,  $\cos\theta_{\mu}$ is the cosine of zenith angle of the muon, $\Phi_{\nu_{\mu}}$ is the muon  neutrino flux, $P(\nu_{\mu}\rightarrow \nu_{\mu})$ is the muon neutrino survival probability,  $\sigma_{\nu_{\mu}}(E_{\nu},E_{\mu})$ is the differential charged current muon-neutrino cross-section.  For  the integration boundaries, $E^{\nu}_{0}$ and $E^{\nu}_{f}$ the initial and final neutrino energies; $\cos\theta_{\mu_{0}}$ and $\cos \theta_{\mu_{f}}$ the bin of zenith angle distribution of Super-Kamiokande experiment, in this we use 10 equal bins of  muon zenith angle between  $\cos\theta_{\mu}={-1,1}$.   We compute this expression for the 2 different types of 
events in SK experiment:  the so-called {\it Sub-GeV} data set and {\it Mult-GeV} data set. They correspond respectively to the intervals of $(E^{{\rm min}}_{\mu},E^{{\rm max}}_{\mu})=(0.2$~GeV, $1.2$~GeV)  and $(E^{{\rm min}}_{\mu},E^{{\rm max}}_{\mu})=(10.0$~GeV, $100.0$~GeV). We use the kinematics of reaction\footnote{For a good reference on kinematical constraints see~\cite{gol}}, defined by $\theta_z,\theta_{\mu}$ and $E_{\nu}$ variables to set-up the allowed range of muon energy,  given by the function $E_{\mu}(\cos\theta_{z},E,\cos\theta_{\mu})$ and we constrain to be in the  {\it Sub-GeV} and {\it Mult-GeV} energy range. Notice that the zenith angle of leptons it is a function of zenith angle of neutrino, the scattering angle  and the energy of neutrino and this produce  a stronger averaging effect on the original neutrino direction.
\begin{figure}[t!]
\centering
\subfloat[]{
\includegraphics[scale=.3,angle=270]{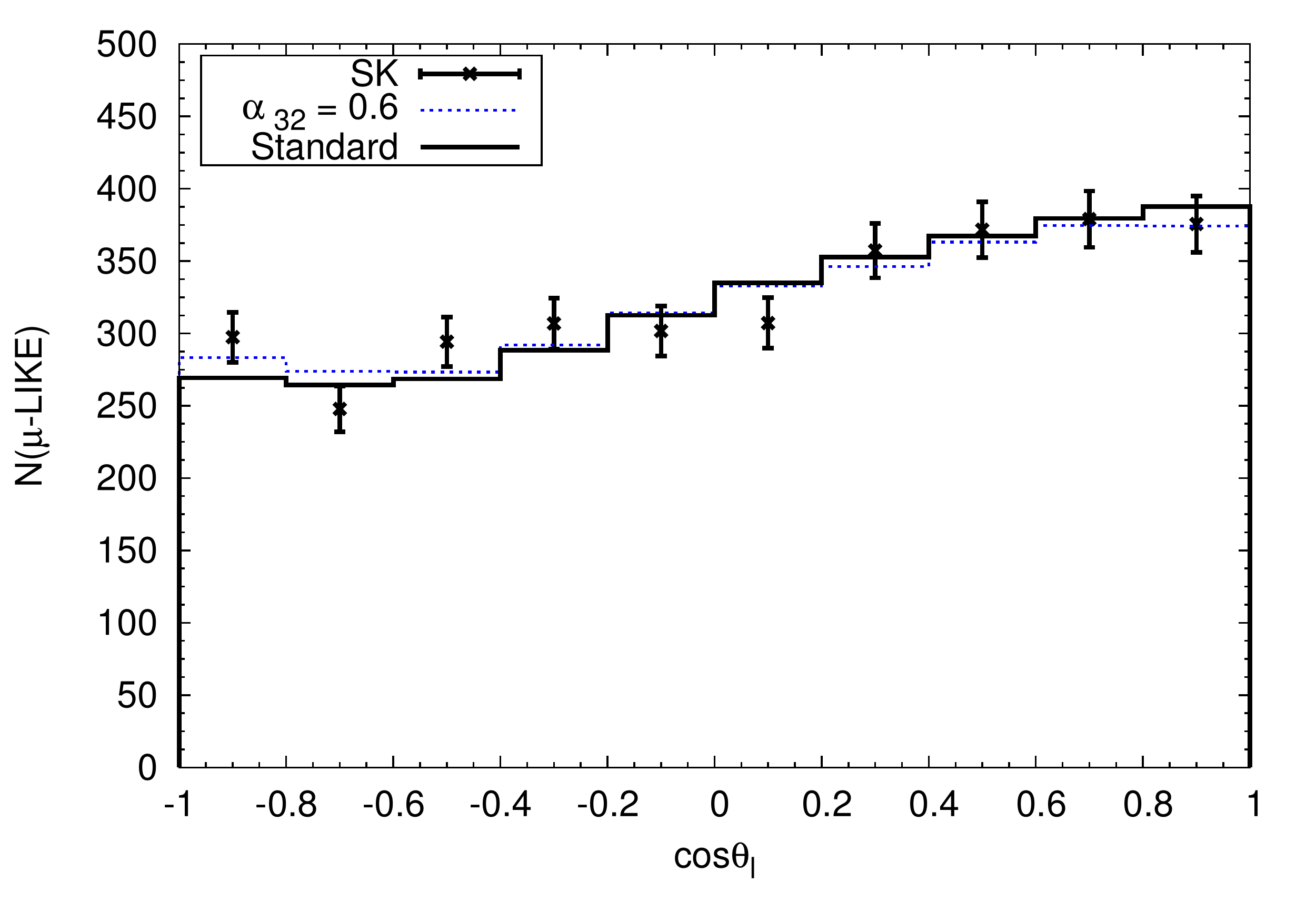}
\label{fig:muonsub}
}
\subfloat[]{
\includegraphics[scale=.3,angle=270]{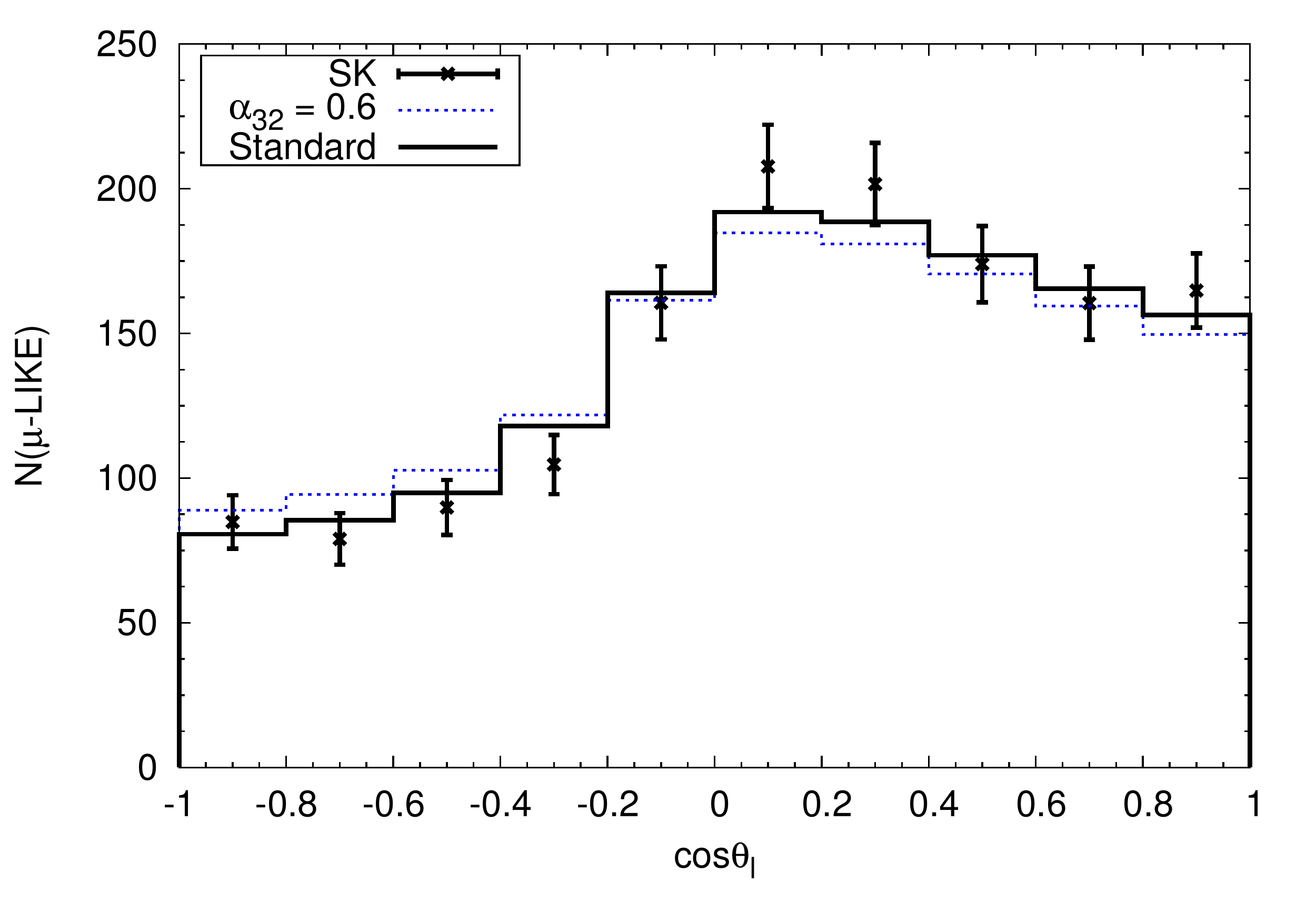}
\label{fig:muonmulti}
}
\quad
\subfloat[]{
\includegraphics[scale=.3,angle=270]{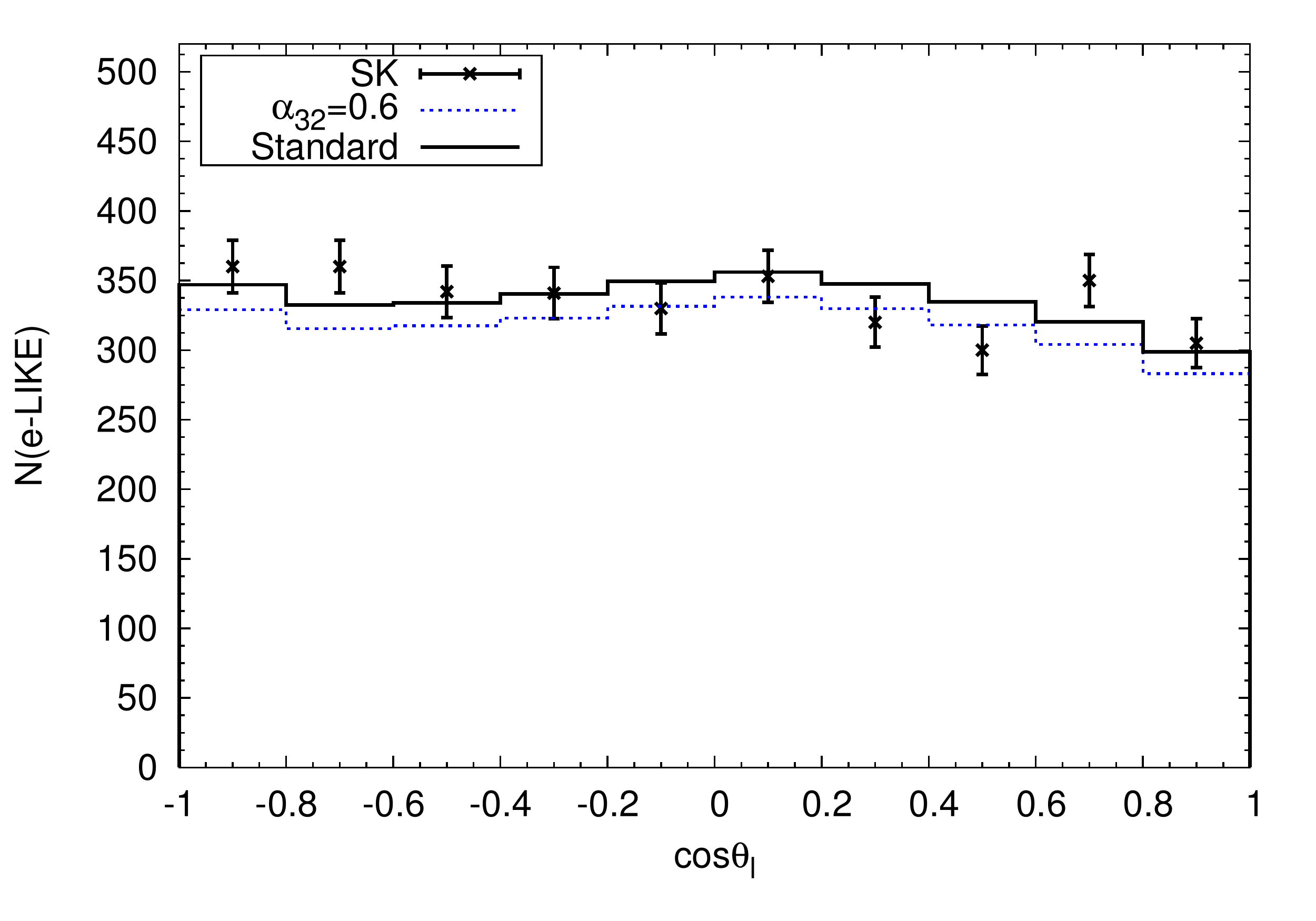}
}
\subfloat[]{
\includegraphics[scale=.3,angle=270]{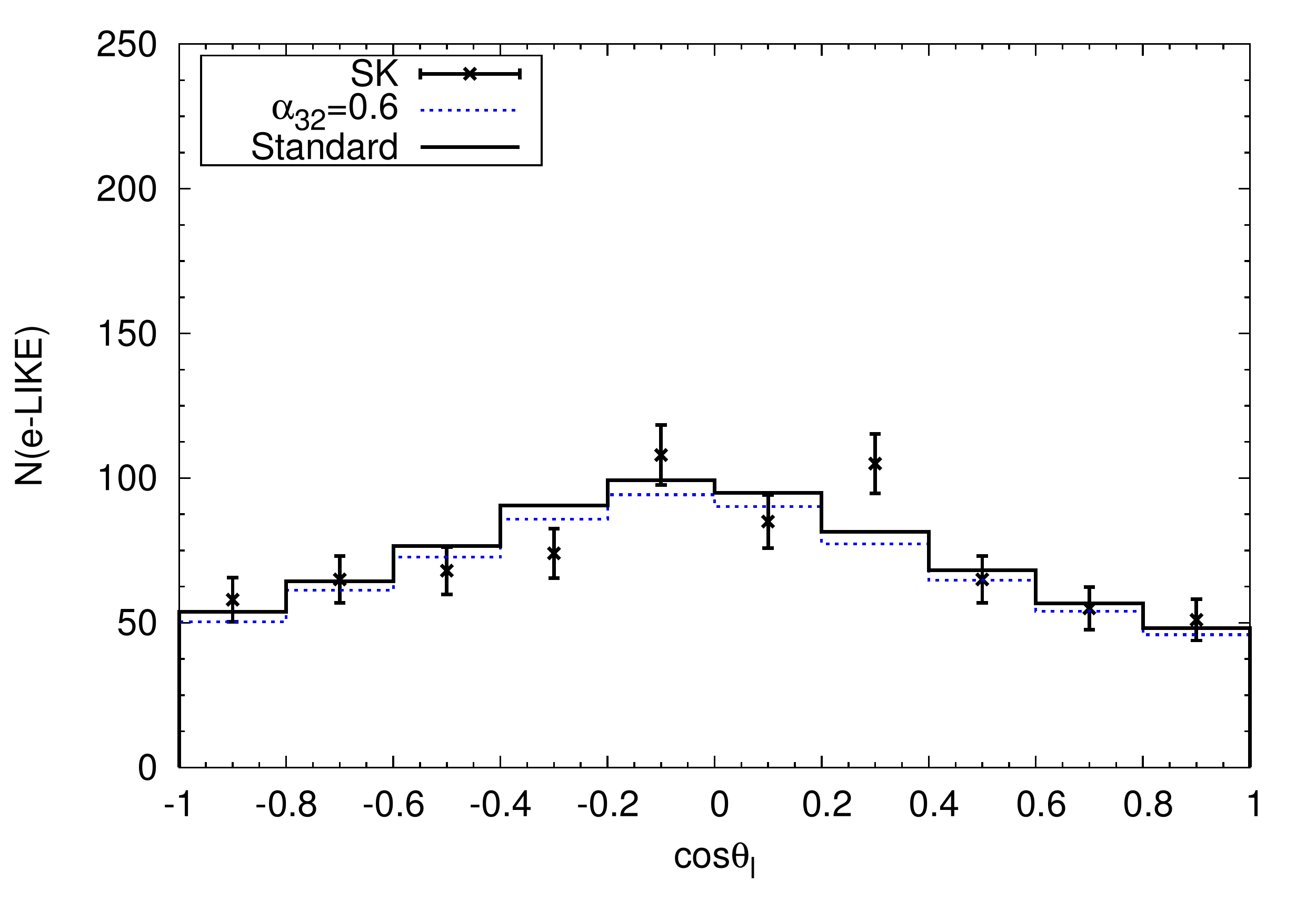}
}
\caption{In sequence, upper (bottom) panels we show the zenith distributions of muon-like (electron-like) events in SK (points). The left (right) panel are for Sub-GeV (Multi-GeV) energy region. Solid line refers to our calculation for the expect number of events for the case of standard neutrino oscillations($\alpha_{32}=0$). The dashed line refers to the case of $\alpha_{32}=0.6$. Points refers to SK data from~\cite{Hosaka:2006zd}.  We assume $\Delta m^2_{32}=2.6\times 10^{-3}$eV$^2$ and maximal mixing angle $\sin^2(\theta_{23})=1/2$}
\label{fig:neve}
\end{figure}
For the electron-like zenith distribution of events in SK we can write
\begin{eqnarray}
N(e)&=&T N_{t}~\int_{E^{\nu}_{0}}^{E^{\nu}_{ f}}dE_{\nu}\int_{0}^{1}dx\int_{-1}^{1}d\cos(\theta_{z})\int_{0}^{2\pi}d \phi_{\nu}\int_{\cos\theta_{e_{0}}}^{\cos\theta_{e_{f}}}d\cos\theta_{e}~  \nonumber \\
&\times& \left \{ \Phi_{\nu_{e}}(E_{\nu},\theta_{\nu},\phi_{\nu})\times \dfrac{d\sigma_{\nu_{e}}(E_{\nu},E_{\mu})}{dE_{\nu}dE_{e}} 
\right.  \nonumber \\
&+ & \left. \Phi_{\bar \nu_{e}}(E_{\nu},\theta_{\nu},\phi_{\nu})\times \dfrac{d\sigma_{ \bar \nu_{e}}(E_{\nu},E_{e})}{dE_{\nu}dE_{\mu}} \right \} \nonumber \\
&\times& \Theta[E_{e}(\cos\theta_{z},E_{\nu},\cos\theta_{e})-E^{\rm min}_{e}]\Theta[E^{\rm max}_{e}-E_{e}(\cos\theta_{z},E_{\nu},\cos\theta_{e})]
 ~,
\label{ne}
\end{eqnarray}
which it is  very similar  to the muon events, with the exception that the electron neutrino did not oscillate, due our initial assumption only have oscillations between muon and tau neutrinos.
The  muon and electron neutrino fluxes are taken from~\cite{Honda:2006qj}. The differential cross section follows~\cite{GonzalezGarcia:1998vk},  whe ere we divide  the cross section in three parts, first the quasi-elastic neutrino scattering  with finite mass corrections~\cite{Strumia:2003zx}; second the one pion contribution and third the DIS contribution~\cite{Paschos:2001np}. Due the Super-Kamiokande did not discriminate between particles and anti-particles, we sum over neutrino and anti-neutrino types.

The muon and electron rate for atmospheric neutrinos have the uncertainties from the prediction of the atmospheric neutrino fluxes, $\Phi_{\nu_{\mu}}$ and $\Phi_{\nu_{e}}$,  computation that can be $ \Delta \left(\Phi_{\nu_{\mu}}\right), \Delta \left(\Phi_{\nu_{e}}\right)=30\%$ and the relative error $\delta \left(\Phi_{\nu_{\mu}}/\Phi_{\nu_{e}}\right)$ of 5\%~\cite{GonzalezGarcia:1998vk}. Because of this error in the absolute normalization we will use as the  physical observable is the zenith distribution of number of events and the absolute value of our prediction will be scaled with the experimental data. Also the smallness of the relative error of muon and electron neutrino fluxes compared the  error in the absolute number 
implies a stronger correlation between the fluxes of muon and electron neutrinos, that we should take into account.

We use this formalism to describe two energy ranges of SK data, the so-called {\it Sub-GeV} and {\it Mult-GeV} data set for electrons and muons, Eq.~(\ref{nmu}) and Eq.~(\ref{ne}).  We compute the muon neutrino survival probability, given in Eq.~(\ref{standardoscillation}), with the oscillation parameters given by  $\Delta m^{2}_{32}=2.6\times 10^{-3}$~eV$^{2}$  and the amplitude $\sin^2(\theta_{23})=0.5$  shown as the  black curve in Fig.~(\ref{fig:muonsub}) and Fig.~(\ref{fig:muonmulti}) respectively for sub-GeV and Multi-GeV sample.  For these parameters we have no oscillation for $\cos\theta_z> -0.1$ and average out oscillation for $\cos\theta_z< -0.6$. Our results  match  the theoretical curves for the number of events for no-oscillation  and standard oscillation of SK experiment. For our computation with MaVaN's probability, we use the numerical solution of Eq.~(\ref{eqevo-drg}) using the matter profile given in Eq.~(\ref{dmevo1}) and we compute the rate for electron and muon (given by Eq.~(\ref{nmu}) and Eq.~(\ref{ne})), the result is the  dashed  curve in  Fig.~(\ref{fig:muonsub}) and Fig.~(\ref{fig:muonmulti}). The effect of non-zero MaVaN, using the same oscillation parameters and the MaVaN parameter $\alpha_{32}=0.6$  it to distort the muon distribution  making a small oscillation for  cosine of lepton zenith $\cos\theta_z>0$, coming from neutrino zenith angle $\cos\theta_{\nu}<0$,  due the averaging effect mentioned after the Eq.~(\ref{nmu}) ;  and suppressing the averaging of neutrino oscillations  for $\cos\theta_z< -0.6$, both behavior are disfavored by the SK data and from this we expect  to have a constrain in the MaVaN parameter from this data.

In atmospheric neutrinos, neutrino can came from different directions and they probe different medium density making the mass difference $\Delta m^2_{\rm MaVaN}$ to be different in each point of the travel. On the other hand, for MINOS, the neutrino travel only cross the upper crust, and we can assume that density along this short chord is constant, we will assume  $\rho_{\rm crust}=3.59$~g/cm$^{3}$ from the PREM model.  We can use the results of Section~(\ref{sec:frame}) of MaVaN mechanism for  constant medium density where the muon survival tprobability is given by Eq.~(\ref{case2MaVaN}) with the MaVaN mass difference given by Eq.~(\ref{Eq:effDM}) and the amplitude by Eq.~(\ref{Eq:sineff}).  The interesting is that the functional form of MaVaN oscillation probability is exactly like the standard oscillation probability, with the replacement
 $\Delta m^{2}_{32}\to \Delta m^{2}_{\rm MaVaN}$ and $\sin^2  2\theta_{23} \to \sin^2  2\theta_{\rm MaVaN}$.

\begin{figure}[t!]
\includegraphics[scale=0.4]{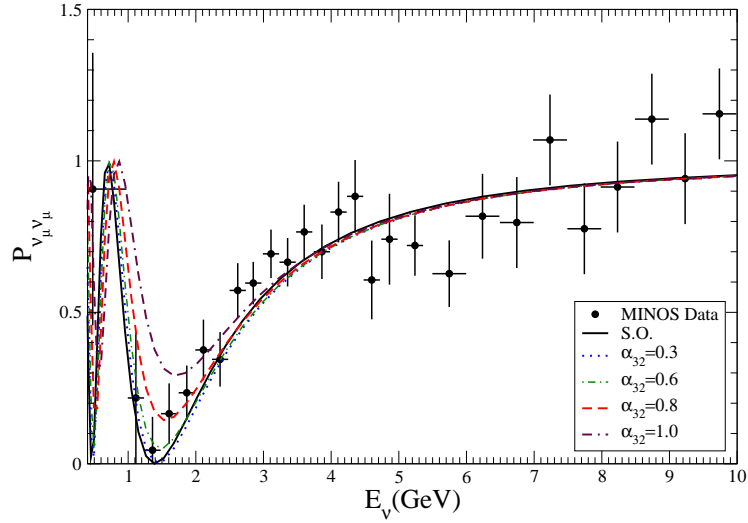}
\caption{Comparison between neutrino oscillations without (S.O.), solid line, and with  MaVaN's, doted and dashed lines. The values of $\alpha_{32}$ are indicated in the figure. Also we assume  $\sin^{2}\theta_{23}=1/2$ and $\Delta m^{2}_{32}=2.5 ~10^{-3}$ eV$^{2}$. Points refers to MINOS data from~\cite{MINOS:beam}.}
\label{minoscte}
\end{figure}
\section{Off-diagonal MaVaN's induced oscillations in MINOS experiment}
\label{sec:minos}
Until now we present a phenomenological framework for off-diagonal MaVaN's mechanism  and show it changes the  $\nu_{\mu}\rightarrow \nu_{\mu}$ atmospheric neutrino oscillation pattern in the SK experiment. However, the SK detector is not the only one sensitive to the neutrino oscillations in the $\nu_{\mu}\rightarrow \nu_{\mu}$ sector and we can use the MINOS data~\cite{Adamson:2013whj}  to constrain the non-diagonal MaVaN parameter.
We show in Fig.~(\ref{minoscte}) the comparison between the standard neutrino oscillation  and the  MaVaN's probability for the setup of MINOS experiment (using the L=L$_{\rm MINOS}$ and the density  $\rho_{\rm crust}=3.59$~g/cm$^{3}$). The oscillations parameters are fixed as  $\Delta m^{2}_{32}=2.5\times 10^{-3}$~eV$^2$ and  maximal mixing $\sin^2 \theta_{23}=1/2$ . When we increase the MaVaN parameter we have for maximal mixing that the MaVaN amplitude get suppressed $\sin^2  2\theta_{\rm MaVaN}\to 0$, as you can see that the minimum, in Fig.~(\ref{minoscte}),   is less deeper for higher values of non-diagonal MaVaN parameter $\alpha_{32}$. At same that the oscillation phase,  $\Delta m^{2}_{\rm MaVaN}$ is bigger for higher values of  $\alpha_{32}$, that it implies that the minimum should be for higher values of energy. We also show for comparison  in Fig.~(\ref{minoscte}) the ratio of experimental number of events over the theoretical prediction without oscillation as data points  to emphasize that when we increase the value of non-diagonal MaVaN parameter we became further apart from the experimental data.

\begin{figure}[t!]
\includegraphics[scale=0.4]{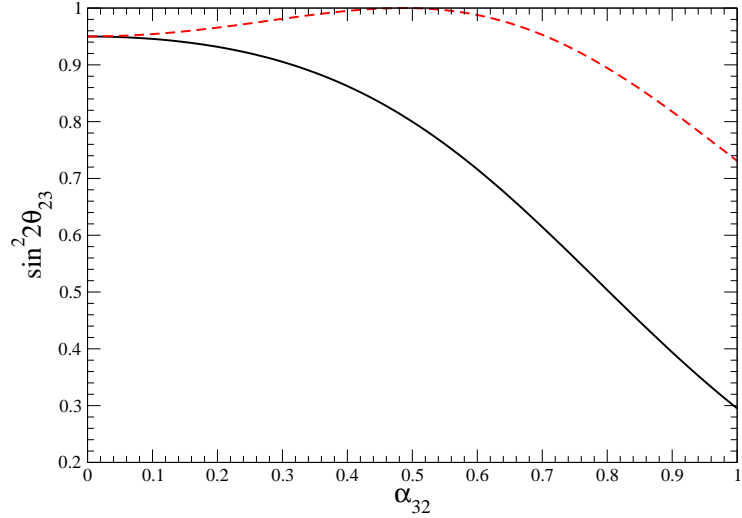}
\caption{The values of  the vacuum mixing angle and $\alpha_{32}$ parameter that  can made a fixed valur for the MaVan amplitude $\sin^2  2\theta_{{\rm MaVaN}}=0.94$}
\label{sineff-ii}
\end{figure}

 So far we have shown examples of MaVaN effects  for maximal vacuum mixing,  $\sin^2 \theta_{23}=1/2$ , e.g.  in Fig.~(\ref{fig:pmumu-cz},\ref{fig:pmumu_alpha0.1},\ref{fig:neve},\ref{minoscte}). A subtle effect can happen for the constant medium density case for larger values of MaVaN parameter, $\alpha_{32}$, in this case we have that the MaVaN amplitude $\sin^2  2\theta_{\rm MaVaN}\to \cos^2  2\theta_{23}$ as commented in Section~(\ref{Sec:model}). For small vacuum mixing angles $\theta_{23}$ this implies that the amplitude is enlarged compared without MaVaN and for larger  vacuum mixing angles $\theta_{23}$ the situation is the opposite, and you have suppression of oscillation. To give a example we display in Fig.~(\ref{sineff-ii}), the vacuum mixing angle  $\sin^2  2\theta_{23}$ as a function of MaVaN parameter  $\alpha_{32}$ that gives a fixed value of MaVaN amplitude $\sin^2  2\theta_{{\rm MaVaN}}=0.94$ (in another words the inverse function of Eq.~(\ref{Eq:sineff})). We can see that it have two solutions: one, the black curve, with small  vacuum mixing angle  $\sin^2  2\theta_{23}$ and another solution given by the  dashed red curve with large vacuum mixing angle. This implies that we can have a small mixing angle and large $\alpha_{23}$ parameter or with can begin a large mixing angle and small parameter $\alpha_{23}$ and both give a effective large MaVaN amplitude. For $\alpha_{23}\rightarrow  \infty $, both curves coincide and we have  full suppression of oscillation amplitude. 
We will see the consequences of subtle effect later in our analysis in Section~(\ref{sec:chi2}).

\section{Analysis of   MINOS and Super-Kamiokande and experiment}
\label{sec:chi2}
Here we give the details of data analysis. First we begin with the analysis of MINOS experiment and later to the Super-Kamiokande analysis for the standard vacuum oscillation scenario and for the MaVaN scenario.

\begin{figure}[t!]
\includegraphics[scale=0.4]{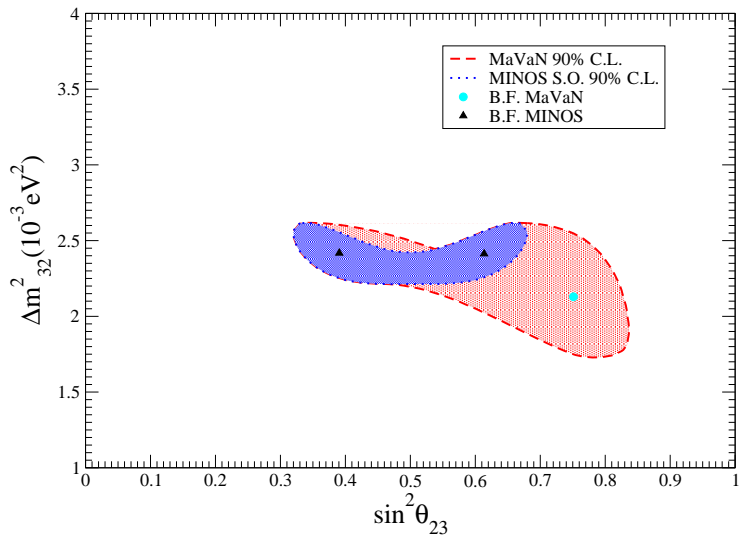}
\caption{Comparison of allowed region at $90\%$ C.L. for oscillations parameters $(\Delta m^{2}_{32},sin^{2}\theta_{23})$ without (S.O), dotted blue  line, and with  MaVaN's,  dashed red line for the MINOS analysis.}
\label{minosonly-regi}
\end{figure}

The  MINOS experiment made a likelihood analysis of theirs data for  {\it neutrinos and anti-neutrinos}  for standard neutrino oscillation  and produces a table with the values of    $(\sin^2 2\theta_{23},\Delta m_{32}^2,\Delta \log L)$, where $\Delta \log L=\log L/L_0$, where $L$ is the likelihood value and $L_0$ is the likelihood for the best fit. This table is  publicly available in Ref.~\cite{minos-chi2}.  We can translated the likelihood language into $\chi^2$  language using $\Delta \chi^2\equiv \chi^2-\chi^2_{\rm min}=2 \Delta \log L $. For the MaVaN analysis, we can use the property that the MaVaN probability have the same functional form of standard oscillation, as discussed in Section~(\ref{sec:minos}), any function of  probability also should have similar behavior. Therefore the $\chi^2_{\rm S.O. MINOS \, analitic}(\sin^2 2\theta_{23},\Delta m_{32}^2)$ function given in  Ref.~\cite{minos-chi2}
as a function of $\sin^2 2\theta_{23}$ and $\Delta m_{32}^2$, should be equal to MaVaN $\chi^2_{\rm MaVaN. MINOS \, analitic}( \sin^2  2\theta_{\rm MaVaN},\Delta m^{2}_{\rm MaVaN})$  as a function of the $\sin^2  2\theta_{\rm MaVaN}$ and $\Delta m^{2}_{\rm MaVaN}$ parameters. Numerically we have
\begin{eqnarray} 
\chi^2_{\rm MaVaN. MINOS \, analitic}( \sin^2  2\theta_{\rm MaVaN},\Delta m^{2}_{\rm MaVaN})=\chi^2_{\rm S.O. MINOS \, analitic}(\sin^2 2\theta_{23},\Delta m_{32}^2)
\label{chi2-mavaniii}
\end{eqnarray}
where for MINOS experiment we can use the expression for MaVaN amplitude $\sin^2  2\theta_{\rm MaVaN}$ and mass difference $\Delta m^{2}_{\rm MaVaN}$
respectively Eq.(\ref{Eq:sineff}) and Eq.(\ref{Eq:effDM}) that give the MaVaN parameters as functions of the vacuum oscillation parameters  $\sin^2 2\theta_{23}$ and $\Delta m_{32}^2$.  We use this procedure to get the allowed region for MINOS only, for the standard oscillation scenario and for the MaVaN case. The results is shown in Fig.~(\ref{minosonly-regi}) where the dotted blue curve correspond to standard neutrino oscillations region of $90\%$ of C.L.. As we plot the region of allowed region in the standard scenario  as an  function of $\sin^{2}\theta_{23}$ (and not in function of $\sin^{2}2\theta_{23}$) then two degenerated minimums  (denoted by up triangles) do appear. When we include MaVaNs and minimize with respect to $\alpha_{32}$ we get the  allowed region is given by the red dashed curve which has the  best fit for non-zero $\alpha_{32}$, and for vacuum mixing angles smaller then maximal $\sin^2\theta_{23}=0.8$ and mass differences $\Delta m^{2}=2.2\times 10^{-3}$~eV$^2$.  In the MaVaN's allowed region the mixing parameters can have  larger values of $\sin^2 \theta_{23}$ and smaller values of  $\Delta m^{2}_{32}$ that  are not allowed in standard oscillation scenario.  Smaller values of   $\Delta m^{2}_{32}\sim 2\times 10^{-3}$~eV$^2$ are not allowed in the standard scenario due it implies smaller  oscillation  effect that is in contradiction with the MINOS data, but in MaVaN mechanism, the effective  mass difference $\Delta m^{2}_{\rm MaVaN}$ can be larger then vacuum oscillation mass difference compensating the smaller value of $\Delta m_{32}^2$. Also the value of $\sin^{2}\theta_{23}\sim 0.8 $ in the MaVaN solution, if we are working in standard scenario will implies a smaller oscillation amplitude that also it is not compatible with MINOS data, but also we can allow values  $\sin^{2}\theta_{23}$ far for maximal for non-zero  $\alpha_{32}$.

Now we will work with the analysis of SK data, where we will use the muon and electron data for Sub-GeV and Multi-GeV samples.  The sample of atmospheric neutrino data is specially interesting for the MaVaN oscillation effect because it is composed by events produced by neutrinos traveling in vacuum and in matter from the 
 use the muon and electron data for Sub-GeV and Multi-GeV sample as discussed in Section III. 

To settle the basics of our analysis we should be aware that the predicted atmospheric neutrino flux have a uncertainty of 30 \% in the normalization and also a stronger correlation between the fluxes, the relative error between the muon and electron flux is around 5\%~\cite{Honda:2006qj}.  
As said before, due this normalization error we will test the shape of muon and electron distribution and not the absolute number of events.   We made the following way, we are going to analysis the shape of atmospheric neutrino data and not include the comparison of the absolute value of data. To do this we will  change our theoretical prediction of oscillation $N_{{\rm theo}}^{\gamma} \to \left(N_{{\rm theo }} ^{{\rm renor}}\right)^{\gamma}\equiv  N_{{\rm theo }} \beta_{\gamma}$, where $\beta_{\gamma}$ is the normalization parameter with a error of $\sigma_{\beta_{\gamma}}=30\%$, with $\gamma=e,\mu$ for each flavor. Also we should use a correlation between the electron and muon number of events due the correlation of the neutrino fluxes.  We made two analysis:
\begin{enumerate}
\item  for the standard oscillation scenario to test our ability to reproduce the results of SK analysis for the oscillation parameters  $\Delta m_{32}^2$ and 
$\sin^2 (\theta_{23})$. We reproduce the main characteristics of with   $\Delta m_{32}^2\sim 3\times10^{-3}$~eV$^2$ and near maximal mixing;
\item the MaVaN scenario for non-diagonal parameter $\alpha_{23}$, with $\alpha_{23}\neq 0$, 
\end{enumerate}
 where our goal is  reproduce the angular distribution predicted by SK experiment using  our computation made independently of the SK experiment and from this to do a reliable analysis of MaVaN phenomena.  Our choice of $\chi^2$ function is 
\begin{equation}
\chi ^{2}_{\rm SK}(\Delta m^{2}_{32},\sin^{2}(\theta_{23}),\alpha_{32})= 
\sum\limits_{ij}\left(
\begin{array}{c}
 N^{\rm data}-\beta N^{\rm teo}
\end{array}
\right)_{i} 
\left(\sigma^2\right)^{-1}_{ij}
\left(
\begin{array}{c}
 N^{\rm data}-\beta N^{\rm teo}
\end{array}
\right)_{j} +\dfrac{(\beta-1)^{2}}{\sigma_{\beta}^{2}}
\label{xq2-new}
\end{equation}
where $N^{\rm data}_{i}$ is the number of events in the bin $i$ measured by SK, $N^{\rm teo}_{i}$ is our prediction for the number of events for the bin $i$ that depend on the oscillation model used;  the non-diagonal matrix $\left(\sigma^2\right)^{-1}_{ij}$ such the diagonal entries have error of $30\%$ as said before, and the non-diagonal entries fixed by the correlation between the muon and electron flux~\cite{GonzalezGarcia:1998vk}. The sum is over the 40 bins:  $40=10 \times 2\times 2 $, counting 2  flavors and 2 samples: Sub-GeV or Multi-GeV, bins.
We add the second term in Eq.~(\ref{xq2-new}) to introduce a penalty factor when  $\beta$ assume values to far from $1$ with respect to the error in normalization, ~$\sigma_{\beta}=0.3$. 
\begin{table}[t!]
\begin{center}
\begin{tabular}{cccccc}
\hline
Model&$(\Delta m^2_{32} eV^{2})_{\rm{b.f.}}$ & $(\sin^2 \theta_{23})_{\rm{b.f.}}$ &  ($\alpha_{\rm{32}})_{\rm{b.f.}}$ & $\Delta\chi^2_{\rm{b.f.}}$ \\
\hline
S. O.&$2.42 \times 10^{-3}$    &0.46 (0.54)   & 0.0  & 1.8 \\
MaVaN's&$2.45 \times 10^{-3}$   & 0.46  & 0.28 & 0.0  \\
\end{tabular}
\end{center}
\caption{ Summary of our  $\chi^{2}$ analysis for combined data from Super-Kamiokande and MINOS experiment. The first line corresponds to pure standard neutrino oscillations(S.O.) and the second line corresponds to MaVaN scenario.  In each case are show the best fit (B.F.) values of the parameters.}
\label{Tab:tab01}
\end{table}

 The MINOS experiment test the MaVaN scenario for a constant density that implies that the MaVaN parameters are fixed, and combining with SK experiment that test the MaVaN for variable density made specially adequate the main hypothesis of MaVaN idea, to have a mass difference and the mixing amplitude that depends on the local density.  To achieve this we combine the two analysis,  from MINOS experiment and from SK experiment , we will use the 
$\chi^{2}$ test for both experiments Super-Kamiokande and MINOS,
\begin{equation}
\chi^{2}_{\rm TOT}(\sin^2 \theta_{23},\Delta m_{32}^2,\alpha_{32})=\chi ^{2}_{\rm SK}+\chi ^{2}_{\rm MINOS}
\label{eq:xq2t}
\end{equation}
With this $\chi^2$ with three oscillation parameters, first we analyze the standard oscillation scenario and second the MaVaN scenario for this combined analysis  of MINOS and SK. W  show  the best fit values for both scenarios in Tab.~(\ref{Tab:tab01}).  From this information we can see that there is a  milder improvement of the MaVaN solution over the standard neutrino oscillation for the combining fit.   For the MaVaN analysis of the  combination of MINOS and SK, we have found that the best fit  is for a non-zero value of MaVaN parameter ($\alpha_{\rm{32}})_{\rm{b.f.}}=0.28$ and the mixing parameters
$(\Delta m^2_{32})_{\rm{b.f.}}=2.45 \times 10^{-3}$~eV$^{2}$ and mixing angle $(\sin^2 \theta_{23})_{\rm{b.f.}}=0.46$ and the best fit parameters for standard oscillation are very similar  $(\Delta m^2_{32})_{\rm{b.f.}}=2.42 \times 10^{-3}$~eV$^{2}$ and $(\sin^2 \theta_{23})_{\rm{b.f.}}=0.46(0.54)$   (see Tab.~(\ref{Tab:tab01}). 
\begin{figure}[t!]
\includegraphics[scale=0.7]{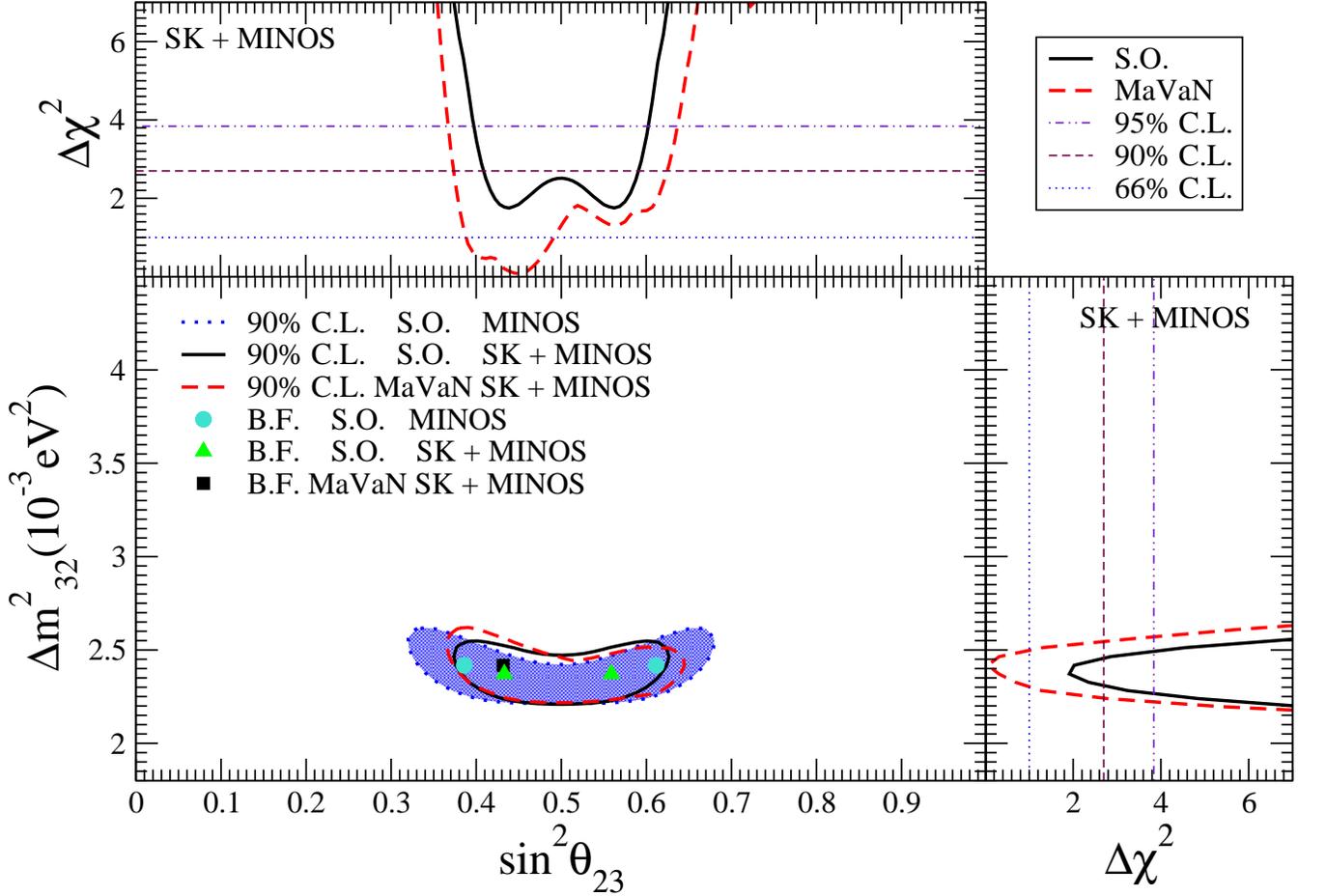}
\caption{In the center panel we show our result of combined  analisys of MINOS and SK experiments for the allowed region of  standard oscillation parameters $(\Delta m^{2}_{32}, ~sin^{2}\theta_{23})$ for the cases with (dashed red curve) and without MaVaN's effect (solid black line) is compared with the result from MINOS Collaboration (dotted blue curve). We also show the values of best fit points respectively in each  analysis.}
\label{fig:deltachi-dmsin-2}
\end{figure}
We shown in the {\it central panel} of Fig.~(\ref{fig:deltachi-dmsin-2}) the allowed region  of $\Delta m^2_{32}$ and  $\sin^2 \theta_{23}$ parameters for 
the  standard oscillation scenario ( S.O. for now) and for the MaVaN scenario for the following cases
\begin{enumerate}
\item first for analysis of the  {\it standard oscillations (S.O.)}  in MINOS experiment alone  shown  the dotted blue curve,
\item second, for the  {\it standard oscillations (S.O.)  analysis}  for the {\it combination} of MINOS and SK experiments shown by  the black curve. We can compare with the MINOS only result the improvement on the determination of range of   $\sin^2 \theta_{23}$ for the combination. The recent values from global fits for these parameters~\cite{GonzalezGarcia:2012sz}  are  included in our $1\sigma$ allowed region for $\Delta m^{2}_{32}$ and $\sin^{2}(2\theta_{23})$,
\item third, for the {\it MaVaN scenario} analysis for the {\it combination} of MINOS and SK experiments shown by dashed red curve,
\end{enumerate}
all plots  shown the 90 \% C.L. allowed region. Comparing the MaVaN  allowed region for MINOS experiment  only shown in Fig.~(\ref{minosonly-regi}) and for the combination of MINOS and SK data  (shown in  central panel of  Fig.~(\ref{fig:deltachi-dmsin-2})) we see that combining the two experiments we constrain more the allowed region of parameters. 

Another good tool to understand our solution is the plot of the projection of $\Delta \chi^2\equiv  \chi^{2}_{\rm TOT}-\chi^2_{\rm b.f.}$ function with respect the 
one of three oscillation parameters, $\sin^2 \theta_{23},\Delta m_{32}^2$ and $\alpha_{32}$. When we show the projection, e. g. for example  $\Delta \chi^2 \times\Delta m_{32}^2$ we have minimized over the other parameters, $\sin^2 \theta_{23}$ and $\alpha_{32}$ and so on. We begin with the  the {\it top panel} of Fig.~(\ref{fig:deltachi-dmsin-2}) we shown the plot of the projection of $\chi^2 \times \sin^2 \theta_{23}$ for the standard oscillation (black curve) and MaVaN (dashed red curve) together  with the 66 \%, 90 \% and 95\% C.L.  respectively $\Delta \chi^2=1.0, 2.70, 3.84$. We can see that the combination  of MINOS and SK suppress the high values of $\sin^2 \theta_{23}>0.7$ that are present in the MINOS only analysis, they appear only at high C.L. (not shown) when  we have  $\Delta \chi^2 > 7$. These high values are only possible in MINOS analysis for $\alpha_{32}>0.7$, however this large value will induce stronger changes in oscillation probabilities for the SK experiment due larger density that the neutrino feels when cross the earth, typically we have  $\rho_{\rm SK}>(2--3) \rho_{\rm MINOS}$. We can notice in Fig.~(\ref{fig:pmumu-cz}) that for  $\alpha_{32}>0.7$ we expect to see a stronger suppression of the muon neutrino oscillation  that will conflict the SK data. 
In the {\it right panel} of Fig.~(\ref{fig:deltachi-dmsin-2}) we shown the plot of the projection of  $\chi^2 \times \Delta m_{32}^2$  for also both standard oscillation and MaVaN scenario respectively by black and dashed red curve. We can see a slightly increase of allowed region for $\Delta m_{32}^2$ when we have a non-zero MaVaN parameter. 
\begin{figure}[t!]
\includegraphics[scale=0.7]{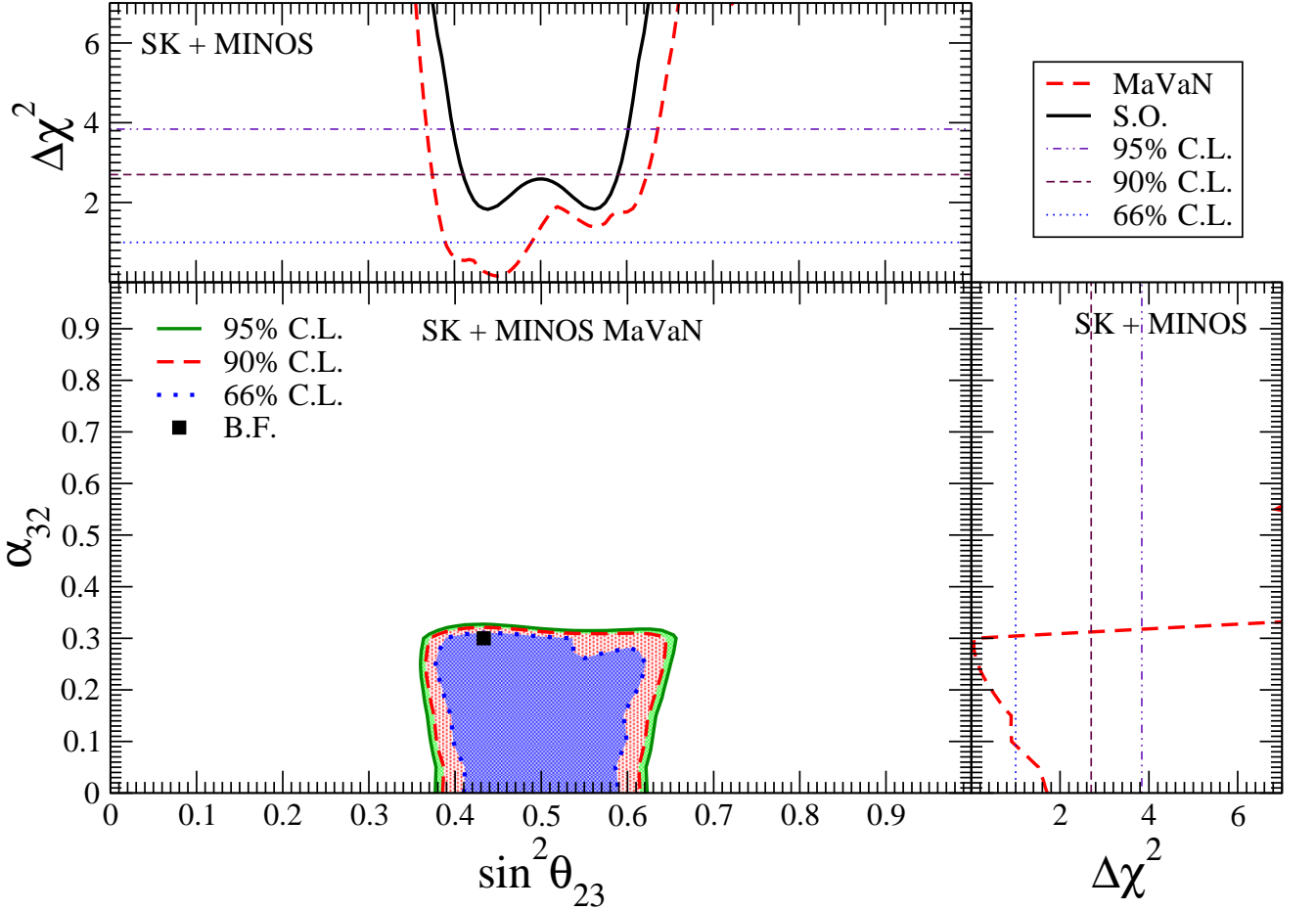}
\caption{In the center pannel we show our result of combined  analisys of MINOS and SK experiments for the allowed region of  parameters $(\sin^{2}\theta_{23},~\alpha_{32})$ for $66\%$(dotted-dashed blue line), $90\%$(dashed red line), and $95\%$(solid green line) C.L.. Black square is  the  best fit point. The auxiliary upper  plot is the same of  Fig.~(\ref{fig:deltachi-dmsin-2}). The left auxiliary plot refers to $\Delta \chi^{2}$ as function of $\alpha_{32}$ and minimized with respect to the other mixing parameters,  $(\sin^{2}\theta_{23}$ and $\Delta m_{32}^2$. }
\label{fig:deltachi_aisin}
\end{figure}
We show in central panel of  Fig.~(\ref{fig:deltachi_aisin}) the allowed region of parameters $(\sin^{2}\theta_{23},~\alpha_{23})$ and minimized with respect to $\Delta m^{2}_{32}$ at  $66\%$, 90 \% and 95 \% C.L.  You can see from this plot that there is a correlation between the $\sin^{2}\theta_{23}$ and $\alpha_{23}$ for  range of values allowed, for the highest  $\alpha_{32}$ value we have the widest range for $\sin^{2}\theta_{23}$. This is left-over of behavior, discussed in Section (\ref{sec:minos}),  that we can have a smaller vacuum mixing amplitude for higher $\alpha_{32}$ or a large vacuum mixing amplitude for smaller $\alpha_{32}$.  In the right panel we show the plot of the projection of $\chi^2\times \alpha_{23}$ where we can see the standard oscillation scenario (when $\alpha_{32}=0$) is compatible at $\Delta \chi^2=1.8$ and the more interesting information from this plot is that there is no allowed region for  $\alpha_{32}>0.32$ at 90 \% C.L. . The solutions for higher $\alpha_{32}>0.5$ present in the MINOS analysis and only appear now for the combined analysis at $>99$\% C.L. for the same reasons discussed in the previous paragraph. 

\begin{figure}[t!]
\includegraphics[scale=0.7]{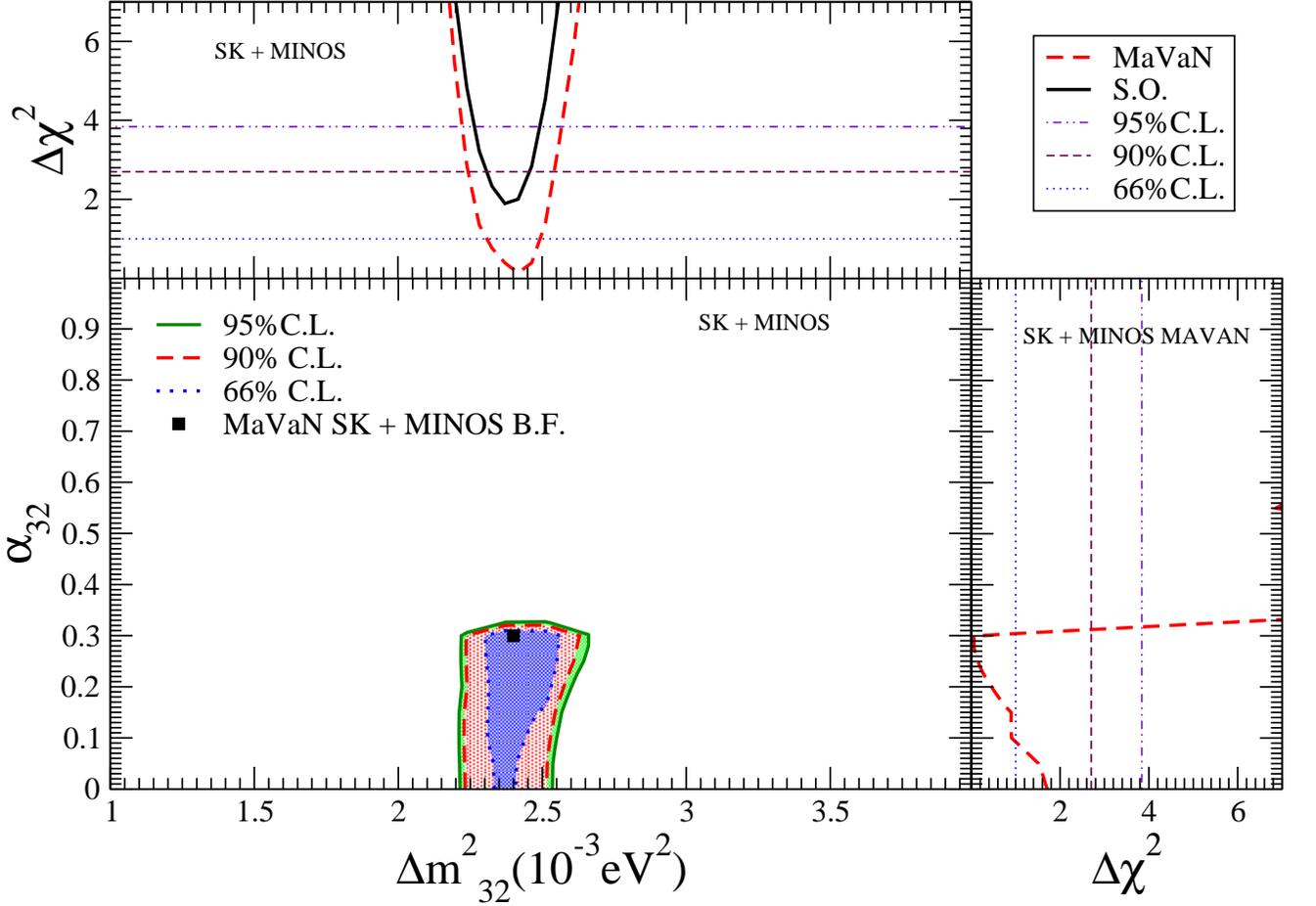}
\caption{In the center pannel we show our result of combined  analisys of MINOS and SK experiments for the allowed region of  parameters $(\Delta m^{2}_{23},~\alpha_{32})$ for $66\%$(dotted-dashed blue line), $90\%$(dashed red line), and $95\%$(solid green line) C.L.. Black square is  the  best fit point. The auxiliary upper(right)  plot refers to $\Delta \chi^{2}$ as function of $\Delta m^{2}_{32}$($\alpha_{32}$) and minimized with respect to other parameters.}
\label{fig:deltachi_dmai}
\end{figure}

 At end, to completeness, in central panel of Fig.~(\ref{fig:deltachi_dmai}) we show the allowed region of parameters as function of $(\Delta m^{2}_{23},~\alpha_{32})$ .  In this central panel, we can see that the correlation between the values of  $\Delta m^{2}_{23}$ and $\alpha_{32}$  are much milder then between the  $(\sin^2 \theta_{23},~\alpha_{32})$  as shown in Fig.~(\ref{fig:deltachi_aisin}). The reason for this is the change in $\Delta m^{2}_{\rm MaVaN}$ due $\alpha_{32}$ it is not so stronger compared with the change in   $(\sin^2 \theta_{\rm  MaVaN}$.

\section{\label{sec:conclusion}Conclusion}

In this work we test  the possibility of neutrinos have their mass  dependent on the medium density. The so-called MaVaN's model includes  this new feature through a new neutrino interaction mediated by a scalar field. We investigate the consequences of such model in the phenomenology of atmospheric neutrino data from Super-Kamiokande experiment and for the data from MINOS experiment using for the first time  to test the existence of non-diagonal MaVaN parameter.

Using the fact the MaVaN for the constant matter density, as it is the case for the MINOS experiment,  have the same functional form of standard oscillation scenario we use the MINOS analysis   for the  standard neutrino oscillation and we extend for the MaVaN scenario. In this analysis larger values of non-diagonal MaVaN parameter, $\alpha_{32} >0.7$, are allowed and we have significant changes in the allowed region for oscillation parameters, $\Delta m^2_{32}$ and  $\sin^{2} (\theta_{23})$.

We compute by ourselves the event rate for Super-Kamiokande experiment,  involves the correct description of  the {\it Sub-GeV} and {\it Multi-GeV} neutrino energy samples and we also use the analysis of the MINOS experiment to made a combined fit of these two experiments.  In MINOS experiment the neutrino crosses only a small chord of the earth and then  the neutrino feels a constant matter density and  for other side  the Super-Kamiokande have neutrinos coming from different directions and therefore feels different medium densities. Then by combining the two data, we can test the essence of MaVaN hypothesis, the mass dependence on the medium density.  We have found that the best fit values for $\Delta m^2_{32}=2.45\times 10^{-3}$~eV$^2$, $\sin^{2} (\theta_{23})=0.42$ and MaVaN parameter $\alpha_{32}=0.28$ and the best fit values for the standard oscillation give similar values(see Table~(\ref{Tab:tab01})). 

The allowed region for the oscillation parameters,  $\Delta m^2_{32}$ and  $\sin^{2} (\theta_{23})$ is shown in Fig~(\ref{fig:deltachi-dmsin-2}) that shows that the allowed region is very stable with the addition of MaVaN scenario.  Although the best fit is for non-zero MaVaN parameter  we have all values of MaVaN parameter  $\alpha_{32}>0.31$ are ruled out at 90\% C.L that allow us to conclude that the non-diagonal MaVaN coupling should be give smaller contribuition to neutrino oscillation.

\begin{acknowledgments} 
The work of D.~R. G. is supported by CAPES and CNPQ. O.~L.~G.~P. thanks the ICTP and the financial support from the  funding grant  2012/16389-1, S\~ao Paulo Research Foundation (FAPESP). 
\end{acknowledgments}

\end{document}